\documentclass[10pt]{article}
\usepackage{amsmath,bm,color,hyperref}
\usepackage[ruled,noend,linesnumbered,figure]{algorithm2e}
\usepackage{graphicx}

\newcommand{\ds}{\displaystyle}

\author{Peter Koval\footnote{koval.peter@gmail.com, CNRS, HiePACS project, LaBRI}, 
        Dietrich Foerster\footnote{d.foerster@cpmoh.u-bordeaux1.fr, CPMOH, University of Bordeaux 1}, and 
        Olivier Coulaud\footnote{olivier.coulaud@inria.fr, INRIA SUD OUEST,  HiePACS project}}

\title{A Parallel Iterative Method for Computing\\ Molecular Absorption Spectra}

\begin{document}

\maketitle

\begin{abstract}
We describe a fast parallel iterative method for computing molecular absorption spectra within
TDDFT linear response and using the LCAO method. We use a local basis
of ``dominant products'' to parametrize the space of orbital products that
occur in the LCAO approach. In this basis, the dynamical polarizability is computed iteratively
within an appropriate Krylov subspace. The iterative procedure uses a
a matrix-free GMRES method to determine the (interacting) density response. 
The resulting code is about one order of magnitude faster than our previous
full-matrix method. This acceleration makes the speed of our TDDFT code comparable
with codes based on Casida's equation. The implementation of our method uses hybrid
MPI and OpenMP parallelization in which load balancing and memory access are optimized.
To validate our approach and to establish benchmarks, we compute spectra of large
molecules on various types of parallel machines.

The methods developed here are fairly general and we believe they will find useful
applications in molecular physics/chemistry, even for problems that are beyond TDDFT,
such as organic semiconductors, particularly in photovoltaics.

\textbf{The manuscript is submitted to Journal of Chemical Theory and Computation, 28.05.2010.}
\end{abstract}

\tableofcontents

\section{Introduction}

The standard way to investigate the electronic structure of matter is
by measuring its response to external electromagnetic fields.
To describe the electronic response of molecules, one may use time-dependent
density functional theory (TDDFT) \cite{TDDFT}.  
TDDFT has been particularly successful in the calculation of absorption
spectra of finite systems such as atoms, molecules and clusters \cite{TDDFT,DFT} and for
such systems it remains the computationally cheapest \textit{ab-initio} approach without
any empirical parameters.

In the framework of TDDFT \cite{Runge-Gross:1984}, time-dependent Kohn-Sham like equations
replace the Kohn-Sham equations of the static density-functional theory (DFT). Although
these equations can be applied to very general situations \cite{TDDFT-Strong-Fields}, we
will restrict ourselves to the case where the external interaction with light is small.
This condition is satisfied in most practical applications of spectroscopy
and can be treated by the linear response approximation.

TDDFT focuses on the dependent electron density $n(\bm{r},t)$. One assumes 
the existence of an artificial time dependent potential 
$V_{\mathrm{KS}}(\bm{r},t)$ in which (equally artificial) non interacting reference electrons
acquire exactly the same time dependent density $n(\bm{r},t)$ as the interacting electrons
under study. The artificial time dependent potential
$V_{\mathrm{KS}}(\bm{r},t)$ is related to
the true external potential $V_{\mathrm{ext}}(\bm{r},t)$
by the following equation
\begin{equation*}
V_{\mathrm{KS}}(\bm{r},t)=V_{\mathrm{ext}}(\bm{r},t)+V_{\mathrm{H}}(\bm{r},t)+
V_{\mathrm{xc}}(\bm{r},t).
\end{equation*}%
Here $V_{\mathrm{H}}(\bm{r},t)$ is the Coulomb potential of the electronic
density $n(\bm{r},t)$ and $V_{\mathrm{xc}}(\bm{r},t)$  the
exchange-correlation potential. 
The exchange-correlation potential absorbs all the non trivial dynamics
and in practice it is usually taken from numerical studies of the interacting homogeneous
electron gas.

The above time dependent extension of the Kohn-Sham equation was used by
Gross \textit{et al} \cite{GrossPetersilka} to find the dynamical linear response 
function $\chi =\frac{\delta n(\bm{r},t)}{\delta V_{\mathrm{ext}}(\bm{r}',t')}$ 
of an interacting electron gas, a response function that expresses
the change of the electron density $\delta n(\bm{r},t)$ upon an external perturbation 
$\delta V_{\mathrm{ext}}(\bm{r}',t')$.
From the change $\delta n(\bm{r},t)$ of the electron density one may calculate
the polarization $\delta \bm{P}=\int\, \delta n(\bm{r},t)\, \bm{r}\,d^{3}r$ 
that is induced by an external field $\delta V_{\mathrm{ext}}(\bm{r},t)$.
The imaginary part of its Fourier transform $\delta \bm{P}(\omega)$ provides us
with the absorption coefficient and the poles of its Fourier transform provide 
information on electronic transitions.

Most practical implementations of TDDFT in the regime of linear response
are based on Casida's equations \cite{Casida,Casida:2009}.
Casida has derived a Hamiltonian in the space of particle-hole excitations,
the eigenstates of which correspond to the
poles of the interacting response function $\chi (\bm{r},\bm{r}^{\prime },\omega )$.
Although Casida's approach has an enormous impact in chemistry \cite{Casida:2009},
it is computationally demanding for large molecules. This is so because Casida's
Hamiltonian acts in the space of particle-hole excitations, the number of which
increases as the square of the number of atoms.

Alternatively, one may also solve TDDFT linear response by iterative methods.
A first example of an iterative method for computing molecular spectra is the direct calculation
of the density change $\delta n(\bm{r},\omega )$ in a modified Sternheimer 
approach \cite{TDDFT,Casida:2009}. 
In this scheme one determines the variation of the Kohn-Sham orbitals
$\delta \psi(\bm{r},t)$ due to an external potential without using any response functions.
By contrast, use of the response functions allowed van Gisbergen \textit{et al}
\cite{vanGisbergen-Baerends:2000,Boeij:2008} to develop a self-consistent iterative
procedure for the variation of the density $\delta n(\bm{r},t)$ without invoking
the variation of the molecular orbitals. There exists also an iterative approach based on
the density matrix  \cite{Rocca-etal:2008,Rocca_thesis}. This approach is quite different 
from ours, but its excellent results, obtained with a plane-wave basis, serve as a useful test
of our LCAO based method. 

Over the last few years, the authors of the present paper developed 
and applied a new basis in the space of products that appear in
the application of the LCAO method to excited states \cite{DF:2008,DF:2009,PK-DF-OC:2010}.
This methodological improvement allows for a simplified iterative approach
for computing molecular spectra and the present paper
describes this approach. We believe that the methods developed here will be useful in molecular physics,
not only in the context of TDDFT, but also for systems such as large organic molecules that,
because of their excitonic features, require methods beyond ordinary TDDFT.  

This paper is organized as follows. In section~\ref{s:theory}, we briefly recall the
TDDFT linear response equations. In section~\ref{s:dominant-products}, we introduce a basis in
the space of products of atomic orbitals. In section~\ref{s:iter}, we describe our iterative
method of computing the polarizability and in section~\ref{s:kernels}, we explain how
the interaction kernels are computed. Section \ref{s:parallelization} describes the parallelization
of our iterative method in both multi-thread and multi-process modes. In section~\ref{s:results},
we present results and benchmarks of the parallel implementation of our code.
We conclude in section \ref{s:conclusion}.

\section{Brief review of linear response in TDDFT}
\label{s:theory}

To find the equations of TDDFT linear response, one starts from the time
dependent extension of the Kohn-Sham equation that was already mentioned in
the introduction and which we rewrite as follows
\begin{eqnarray*}
V_{\mathrm{KS}}([n],\bm{r},t) &=&V_{\mathrm{ext}}([n],\bm{r},t)+
V_{\mathrm{H}}([n],\bm{r},t)+V_{\mathrm{xc}}([n],\bm{r},t).
\end{eqnarray*}%
The notation $[n]$ indicates that the potentials
 $V_{\mathrm{KS}}$, $V_{\mathrm{H}}$ and $V_{\mathrm{xc}}$
in this equation depend on the distribution of electronic charge 
$n(\bm{r},t)$ in all of space and at all times. To find out
how the terms of this equation respond to a small variation 
$\delta n(\bm{r},t)$ of the electron density, we take their
variational derivative with respect to the electron density
\begin{equation*}
\frac{\delta V_{\mathrm{KS}}([n],\bm{r},t)}{\delta n(\bm{r}',t')} =%
\frac{\delta V_{\mathrm{ext}}([n],\bm{r},t)}{\delta n(\bm{r}',t')}%
+\frac{\delta }{\delta n(\bm{r}',t')}\left[ V_{\mathrm{H}}([n],\bm{r},t)+
V_{\mathrm{xc}}([n],\bm{r},t)\right].
\end{equation*}%
The important point to notice here is that
$\displaystyle\frac{\delta V_{\mathrm{KS}}}{\delta n}$ and $\displaystyle\frac{\delta V_{\mathrm{ext}}}{\delta n}$ are 
inverses of $\displaystyle\chi_0=\frac{\delta n}{\delta V_{\mathrm{KS}}}$ and
$\displaystyle\chi=\frac{\delta n}{\delta V_{\mathrm{ext}}}$ 
that represent the response of the density of free and interacting electrons to, respectively,
variations of the potentials $V_{\mathrm{KS}}$ and $V_{\mathrm{ext}}$.
As the response $\chi_0$ of free electrons is known and since we wish to find the response  
$\chi$ of interacting electrons, we rewrite the previous equation compactly
as follows
\begin{equation}
\chi^{-1}=\chi_{0}^{-1}-f_{\mathrm{Hxc}}.  \label{Gross_equation}
\end{equation}
Here an interaction kernel $\ds f_{\mathrm{Hxc}} =\frac{\delta }{\delta n}\left[ V_{\mathrm{H}}+
V_{\mathrm{xc}}\right]$ is introduced.
Equation (\ref{Gross_equation}) has the form of a Dyson equation that is familiar
from many body perturbation theory \cite{Fetter-Walecka:1971}.

To make use of equation (\ref{Gross_equation}), it remains to specify $\chi_{0}$ and $f_{\mathrm{Hxc}}$.
The Kohn-Sham response function $\chi_{0}$ can be computed in terms of orbitals and eigenenergies 
\cite{TDDFT} as will be discussed later in this section (see equation (\ref{response-eig}) below).
For the exchange-correlation potential $V_{\mathrm{xc}}([n],\bm{r},t)$, we use the adiabatic local density
approximation (ALDA) that is local in both time and space,
$V_{\mathrm{xc}}([n],\bm{r},t) = V_{\mathrm{xc}}(n(\bm{r},t))$, therefore
the interaction kernel $f_{\mathrm{Hxc}}$ will not depend on 
frequency 
\begin{equation*}
f_{\mathrm{Hxc}} =\frac{1}{|\bm{r}-\bm{r}'|}+
\delta (\bm{r}-\bm{r}')\frac{dV_{\mathrm{xc}}}{dn}.
\end{equation*}
In principle one could determine the interacting response function 
$\chi (\bm{r},t,\bm{r}',t')$ from equation (\ref{Gross_equation}),
determine the variation of density $\delta n(\bm{r},t)=\int \chi (\bm{r},t,
\bm{r}',t')\delta V_{\mathrm{ext}}(\bm{r}',t')d^3r' dt'$ due to a variation 
of the external potential $\delta V_{\mathrm{ext}}$ and find the 
observable polarization $\delta \bm{P}=\int \delta n(\bm{r},t)\bm{r}\,d^{3}r$.
To realize these operations in practice, we use a basis of localized functions
to represent the space dependence of the functions $\chi_0(\bm{r},\bm{r}',\omega)$
and $f_{\mathrm{Hxc}}(\bm{r}, \bm{r}')$.

In order to introduce such a basis into the Petersilka--Gossmann--Gross equation (\ref{Gross_equation}),
we eliminate the inversions in this equation and transform to the frequency domain
\begin{equation}
\chi(\bm{r},\bm{r}', \omega) = \chi_0(\bm{r},\bm{r}', \omega ) + 
\int d^3r_{1} d^3r_{2}\,\chi_{0}(\bm{r},\bm{r}_1,\omega ) f_{\mathrm{Hxc}}(\bm{r}_1,\bm{r}_2)
\chi(\bm{r}_2,\bm{r}',\omega).
\label{dyson-integral-time}
\end{equation}
The density response function of free electrons $\chi_0(\bm{r},\bm{r}',\omega)$
can be expressed \cite{TDDFT} in terms of molecular Kohn-Sham orbitals as follows%
\begin{equation}
\chi_{0}(\bm{r},\bm{r}',\omega )=\sum_{E,F}(n_{F}-n_{E})\frac{\psi
_{E}(\bm{r})\psi _{F}(\bm{r})\psi _{F}(\bm{r}')\psi _{E}(\bm{r}')}{
\omega -(E-F)+\mathrm{i}\varepsilon }.  \label{response-eig}
\end{equation}
Here $n_{E}$ and $\psi_{E}(\bm{r})$ are occupation factors and Kohn-Sham
eigenstates of eigenenergy $E$, and the constant $\varepsilon $ regularizes
the expression, giving rise to a Lorentzian shape of the resonances. The
eigenenergies $E$ and $F$ are shifted in such a way that occupied and
virtual states have, respectively, negative and positive energies. Only
pairs $E$,$F$ of opposite signs, $E\cdot F<0$, contribute in the summation
as is appropriate for transitions from occupied to empty states.

We express molecular orbitals $\psi_{E}(\bm{r})$ as linear combinations
of atomic orbitals (LCAO)
\begin{equation}
\psi_{E}(\bm{r})=X_{a}^{E}f^{a}(\bm{r}),
\label{lcao}
\end{equation}
where $f^{a}(\bm{r})$ is an atomic orbital. The coefficients $X_{a}^{E}$ are determined by diagonalizing the Kohn-Sham
Hamiltonian that is the output of a prior DFT calculation and we assume these
quantities to be available. For the convenience of the reader, we use Einstein's
convention of summing over repeated indices.

\section{Treatment of excited states within LCAO}
\label{s:dominant-products}

The LCAO method was developed in the early days of quantum mechanics to express
molecular orbitals as linear combinations of atomic orbitals. When inserting the LCAO ansatz
(\ref{lcao}) into equation (\ref{response-eig}) to describe the density response,
one encounters products of localized functions $f^{a}(\bm{r}) f^{b}(\bm{r})$
--- a set of quantities that are known to be linearly dependent.

There is an extensive literature \cite{Beebe-etal:1977,Boys-Shavitt:1959,Skylaris-etal:2000,Baerends-etal:1973}
on the linear dependence of products of atomic orbitals.
Baerends uses an auxiliary basis of localized functions to represent the electronic density
\cite{Baerends-etal:1973,Te-Velde-etal:2001}.
His procedure of fitting densities by auxiliary functions is essential both for solving
Casida's equations and in van Gisbergen's iterative approach.

In the alternative approach of Beebe and coauthors \cite{Beebe-etal:1977}, one forms
the overlaps of products $\langle ab|a' b'\rangle$ to
disentangle the linear dependence of the products $f^{a}(r)f^{b}(r)$.
The difficulty with this approach is its lack of locality and the $O(N^4)$
 cost of the construction of the overlaps \cite{Vysotskiy-Cederbaum:2009}.

Our approach is applicable to numerically given atomic orbitals of finite support and, in the special case of products on
coincident atoms, 
it resembles an earlier construction by Aryasetiawan and Gunnarsson in the muffin tin context
\cite{Aryasetiawan-Gunnarsson:1994}. We ensure locality by focusing
on the products of atomic orbitals for each overlapping pair of atoms at a time
\cite{DF:2008,DF:2009}.
Because our construction is local in space, it requires only $O(N)$ operations.
Our procedure removes a substantial part of the linear dependence from the set of products
$\{f^{a}(\bm{r})f^{b}(\bm{r})\}$. As a result, we find a vertex like identity for the 
original products in terms of certain ``dominant products'' $F^{\lambda }(\bm{r})$
\begin{equation}
f^{a}(\bm{r})f^{b}(\bm{r})\sim\sum_{\lambda >\lambda _{\min }}
V_{\lambda}^{ab}F^{\lambda }(\bm{r}).  \label{vertex-ansatz}
\end{equation}%
Here the notation $V_{\lambda }^{ab}$ alludes to the fact that
the vertex $V_{\lambda}^{ab}$ was obtained from an eigenvalue problem for the pair
of atoms that the orbitals $a$, $b$ belong to. The condition
$\lambda >\lambda _{\min }$ says that the functions corresponding to
the (very many) small eigenvalues were discarded. By construction, the vertex
$V_{\lambda}^{ab}$ is non zero only for a few orbitals $a$, $b$ that refer to
the same pair of atoms that $\lambda$ belongs to and, therefore, $V_{\lambda}^{ab}$ is a sparse object.
Empirically, the error in the representation (\ref{vertex-ansatz})  vanishes exponentially fast
\cite{Larrue:2008} with the number of eigenvalues that are retained.
The convergence is fully controlled by choosing the threshold for eigenvalues $\lambda_{\min }$
and from now on we assume equality in relation (\ref{vertex-ansatz}).

We introduce matrix representations of the response functions $\chi_{0}$ 
and $\chi $ in the basis of the dominant products $\{F^{\mu }(\bm{r})\}$ as follows
\begin{equation}
\chi_{0}(\bm{r},\bm{r}^{\prime },\omega )=\sum_{\mu \nu }F^{\mu }(\bm{r})
\chi_{\mu \nu }^{0}(\omega )F^{\nu }(\bm{r}');\ \ \ 
\chi(\bm{r},\bm{r}^{\prime },\omega )=\sum_{\mu \nu }F^{\mu }(\bm{r})
\chi_{\mu \nu }(\omega )F^{\nu }(\bm{r}').
\label{response-domi}
\end{equation}%
For the non interacting response $\chi^0_{\mu\nu}(\omega)$, one has an explicit 
expression 
\begin{equation}
\chi_{\mu \nu }^{0}(\omega )=\sum_{abcd,E,F}\,(n_{F}-n_{E})\frac{%
(X_{a}^{E}V_{\mu }^{ab}X_{b}^{F})(X_{c}^{F}V_{\nu }^{cd}X_{d}^{E})}{\omega
-(E-F)+\mathrm{i}\varepsilon },  \label{response-mat}
\end{equation}
which can be obtained by inserting the LCAO ansatz (\ref{lcao}) and 
the vertex ansatz (\ref{vertex-ansatz}) into equation
(\ref{response-eig}).

We insert the expansions (\ref{response-domi}) into the Dyson
equation (\ref{dyson-integral-time}) and obtain the 
Petersilka-Gossmann-Gross equation in matrix form
\begin{equation}
\chi _{\mu \nu }(\omega )=\chi_{\mu\nu}^{0}(\omega )+
\chi_{\mu \mu'}^{0}(\omega )
f_{\mathrm{Hxc}}^{\mu'\nu'}(\omega )\chi_{\nu'\nu }(\omega),
\label{matrix_representation}
\end{equation}
with the kernel $f_{\mathrm{Hxc}}^{\mu\nu }$ defined as
\begin{equation}
f_{\mathrm{Hxc}}^{\mu\nu }=\int d^3 r \,d^3 r'\,F^{\mu }(\bm{r})
f_{\mathrm{Hxc}}(\bm{r},\bm{r}')F^{\nu }(\bm{r}').  \label{matrix_self_energy}
\end{equation}
The calculation of this matrix will be discussed in section \ref{s:kernels}.

Since molecules are small compared to the wavelength of light, one may use
the dipole approximation and define the polarizability tensor
\begin{equation*}
P_{ik}(\omega )=\int d^3r  d^3r' \bm{r}_{i}\bm{r}'_{k} \chi(\bm{r},\bm{r}', \omega).
\end{equation*}
Moreover, using equation (\ref{matrix_representation}) we find an explicit expression
for the interacting polarizability tensor $P_{ik}(\omega)$ in terms of the known matrices $\chi^0$ and 
$f_{\mathrm{Hxc}}$
\begin{equation}
P_{ik}(\omega ) = d^{\mu}_i
\left( \frac{1}{1-\chi^{0}(\omega )f_{\mathrm{Hxc}} }\right)_{\mu \nu}
\chi^0_{\nu\nu'}(\omega) d^{\nu'}_k,
\label{eqpolarizability}
\end{equation}%
where the dipole moment has been introduced $d^{\mu}_i=\int F^{\mu }(\bm{r})\bm{r}_i\, d^3r$.
Our iterative procedure for computing molecular absorption spectra is based on equation (\ref{eqpolarizability}).
Below, we will omit Cartesian indices $i$ and $k$ because we compute tensor components $P_{ik}(\omega)$
independently of each other.

\section{An iterative method for the calculation of the dynamical polarizability}
\label{s:iter}

In equation (\ref{eqpolarizability}) the polarizability $P(\omega )$ is expressed
as a certain \textit{average} of the inverse of the matrix
$A_{\mu}^{\nu}\equiv \delta^{\nu }_{\mu } - \chi^{0}_{\mu\nu'}(\omega)f^{\nu'\nu}_{\mathrm{Hxc}}$.
A direct inversion of the matrix $A^{\mu}_{\nu}$ is straightforward at this point, but it is
computationally expensive and of unnecessary (machine) precision.
Fortunately, iterative methods are available that provide the desired result
at much lower computational cost and with a sufficient precision.

In fact, we already used an iterative biorthogonal Lanczos method to create an
approximate representation of the matrix $A^{\mu}_{\nu}$,
which can be easily inverted within the Krylov subspace under consideration \cite{DF:2009}.
In this work, however, we use a simpler approach of better computational stability to compute the
polarizability (\ref{eqpolarizability}). First, we calculate an auxiliary vector 
$X=A^{-1} \chi^0 d$ by solving a linear system of equations $AX=\chi^{0}d$ with an iterative method of
the Krylov type. Second, we compute the dynamical polarizability as a scalar product $P=d^{\mu}\, X_{\mu}$.
In this way we avoid the construction of the full and computationally expensive response matrix $\chi_0$.
Below, we give details on this iterative procedure. 

\subsection{The GMRES method for the iterative solution of a linear system of equations}
\label{ss:gmres}

As we explained above, the polarizability $P(\omega)$ can be computed separately for each
frequency, by solving the system of linear equations
\begin{equation}
\left(1 - \chi^{0}(\omega)f_{\mathrm{Hxc}}\right)X(\omega) =\chi^{0}(\omega) d,
\label{sle}
\end{equation}
and by computing the dynamical polarizability as a scalar product $P(\omega)=d^{\mu} X_{\mu}(\omega)$.

We apply a generalized minimal residual method (GMRES) \cite{Saad,gmres-luc} to solve
the linear system of equations (\ref{sle}), which is of the form $AX=b$.
GMRES belongs to the Krylov-type methods
\cite{Saad,Ipsen-Meyer:98} that represent a large matrix $A$ in
an iteratively built up Krylov-type basis $|0\rangle ,|1\rangle \ldots |i\rangle $.
The first vector $|0\rangle $ in the basis is chosen equal to $|b\rangle$,
while further vectors are computed
recursively via  $|i\rangle =A|i-1\rangle$. As the vectors $|i\rangle
=A^{i}|0\rangle $ are not mutually orthogonal, one may enforce their
orthogonality by using the Gram-Schmidt method 
\begin{equation}
|i\rangle =A|i-1\rangle -\sum_{j=0}^{i-1} |j\rangle  \langle j|A|i-1\rangle.
\label{orthogonal-krylov-basis}
\end{equation}%
The orthonormal basis built in this way is used in the GMRES method to
approximately solve the linear system of equations $A|X\rangle=|b\rangle$ by minimizing the
residual $|r\rangle=A|X\rangle-|b\rangle$ within the Krylov-type subspace (\ref{orthogonal-krylov-basis}).
The minimization of the residual occurs when the equation $\sum_{j}\langle i|A|j\rangle
\langle j|x\rangle =\langle i|b\rangle $ is satisfied and this set of equations is
of much smaller size than the original problem. When the solution in the Krylov subspace
$\langle i|x\rangle $ is found, then an approximate solution in the original
space can be computed from $|X\rangle =\sum_{i}|i\rangle \langle i|x\rangle$.

A suitable stopping criterion is essential for our method and we tested
several criteria in order to keep the number of iterations small and achieve
a reliable result at the same time. The conventionally used criterion that $\varepsilon_r=|r|/|b|$
should be small is unreliable 
when the tolerance threshold is comparatively large
($\varepsilon_r\approx 1\%$).
Therefore we suggest an alternative combined criterion.

A natural stopping criterion for an iterative solver of the linear system of equations
$AX=b$ is a condition on the relative error of the solution $\varepsilon_X=|\Delta X|/|X|$. 

In our case, the quantity of interest is the dynamical polarizability $P=\langle d|X\rangle $,
therefore it is meaningful to impose a limit on the relative error of the
polarizability $\varepsilon_P =|\Delta P|/|P|$. Estimations of the errors
$\varepsilon_X$ and $\varepsilon_P$ can be easily obtained
because $|\Delta X\rangle=A^{-1}|r\rangle$ and $\Delta P=\langle d|A^{-1}|r\rangle$.
We estimate $|\Delta X\rangle$ and $\Delta P$ using a matrix norm \cite{templates},
$\Delta X\approx|A^{-1}|| r\rangle$ and 
$\Delta P\approx|A^{-1}| \langle d|r\rangle$. 
We used the Frobenius norm of the Krylov representation of the matrix  $A^{-1}$,
$|A^{-1}|\approx\sqrt{\sum_{ij} |\langle i|A^{-1}| j\rangle|^2}$ because $A$ is a 
non Hermitian matrix.

Both errors $\varepsilon_X$ and $\varepsilon_P$
tend to zero in the limit $X\rightarrow X_{\mathrm{exact}}$, but for a threshold 
in the range of a few percent they behave differently. The condition upon
$\varepsilon_P$ is better tailored to the problem and it saves unnecessary
iterations in many cases. However, in a few cases, the condition upon $\varepsilon _{P}$
fails, while the condition upon $\varepsilon_{X}$ works. Therefore, we use a
condition on $\varepsilon_P$ in ``quick and dirty'' runs and
a combined condition $\varepsilon_P<\varepsilon _{\mathrm{tolerance}}$ 
and $\varepsilon_X<\varepsilon _{\mathrm{tolerance}}$ for reliable results.

In general, iterative methods of the Krylov type involve only matrix--vector
products $A|z\rangle$. For an explicitly given matrix $A$, the operation $|z\rangle \rightarrow A|z\rangle$ requires
$O(N^2)$ operations. Therefore the whole iterative method will scale as $O(N^2 N_{\mathrm{iter}})$,
where $N_{\mathrm{iter}}$ is the number of iterations until convergence. This is better
than direct methods when $N_{\mathrm{iter}}\ll N$, because a matrix--matrix multiplication takes $O(N^3)$ operations.

To avoid matrix multiplications, the application of the matrix $A=1 - \chi^{0}(\omega)f_{\mathrm{Hxc}}$
to a vector $|z\rangle$ is done sequentially by computing first 
$|z'\rangle=f_{\mathrm{Hxc}}|z\rangle$ and then $A|z\rangle = |z\rangle-\chi^0(\omega) |z'\rangle$.
The kernel matrix $f_{\mathrm{Hxc}}$ is computed before the iterative procedure. Because
it is frequency independent, it can be easily stored and reused. By contrast, the response
matrix $\chi^0(\omega)$ is frequency dependent and computationally expensive and its explicit construction should be avoided. 
Therefore, only matrix--vector products $\chi^0(\omega) |z\rangle$ will be
computed as explained below without ever calculating the 
full response matrix $\chi^0(\omega)$.

\subsection{Fast application of the Kohn-Sham response matrix to vectors}
\label{ss:matrix-vector}

In previous papers \cite{DF:2009,PK-DF-OC:2010}, we have described an $O(N^{2})$ construction
of the entire response function $\chi_{\mu\nu}^{0}(\omega)$, but the prefactor was large.
Paradoxically, the Krylov method presented above allows for a much faster computation of
the absorption spectrum, although the cost of matrix--vector products
$\chi^0(\omega) |z\rangle$ scales asymptotically as $O(N^3)$ (see below).

Starting point of our construction of the matrix--vector product $\chi^0(\omega)|z\rangle$
is the expression (\ref{response-mat}) for the Kohn-Sham response matrix in the basis of dominant products
\begin{equation*}
\chi^0_{\mu\nu}(\omega) z^{\nu} = \sum_{abcd,E,F}\,(n_{F}-n_{E})
\frac{(X_{a}^{E}V_{\mu }^{ab}X_{b}^{F})(X_{c}^{F}V_{\nu }^{cd}X_{d}^{E})}{
\omega-(E-F)+\mathrm{i}\varepsilon } z^{\nu}.
\end{equation*}
To compute this matrix--vector product efficiently, 
we decompose its calculation into a sequence of multiplications that 
minimizes the number of arithmetical operations by exploiting the sparsity of the vertex $V_{\mu}^{ab}$.
\begin{figure}[hbt]
\setlength{\fboxrule}{1pt}
$$\chi^0_{\mu\nu} z^{\nu} \sim \fcolorbox{red}{white}{$\sum_{E}X_a^EV_{\mu}^{ab}
\framebox{$\sum_{F}X_b^F\fcolorbox{blue}{white}{$X_c^F \fcolorbox{green}{white}{$
\fcolorbox{red}{white}{$V_{\nu}^{cd}X_d^E$}
\,z^{\nu}$}$}$}$}$$
\caption{A sequence of operations to compute the matrix--vector product.
$\chi^0_{\mu\nu} z^{\nu}$. 
}
\label{f:flops}
\end{figure}
The sequence we chose is graphically represented in
figure~\ref{f:flops}. For clarity, the frequency dependent denominator
is omitted. Boxes represent the products to be performed
at different steps. The innermost box contains a frequency
independent quantity, that is also used in the last step. An algebraic 
representation of the computational steps is given in table \ref{t:flops}.
\begin{table}[htb]
\tabcolsep 1mm 
\centerline{
\begin{tabular}{|cc||c|c|r|}
\hline
Step & Expression & Complexity & Memory & Details of the Complexity \\ \hline
1 & $\alpha^{cE}_{\nu} = V_{\nu}^{cd} X^{E}_d $ & $O(N^2)$ & $O(N^2)$ & $%
N_{occ} n_{orb}^2 N_{prod} $ \\[1ex] 
2 & $\beta^{cE}=\alpha^{cE}_{\nu} v^{\nu} $ & $O(N^2)$ & $O(N^2)$ & $%
N_{\omega} N_{occ} n_{orb} N_{prod}$ \\[1ex] 
3 & $a^{FE}=X^F_{c} \beta^{cE}$ & $O(N^3)$ & $O(N^2)$ & $N_{\omega}
N_{occ} N_{virt} N_{orb}$ \\[1ex] 
4 & $\gamma^{E}_b=\sum_F X^F_{b} a^{FE}$ & $O(N^3)$ & $O(N^2)$ & $%
N_{\omega} N_{occ} N_{virt} N_{orb}$ \\[1ex] 
5 & $n_{\mu}=\sum_E \alpha^{bE}_{\mu} \gamma^{E}_b$ & $O(N^2)$ & $O(N)$ & $%
N_{\omega} N_{occ} n_{orb} N_{prod}$ \\[1ex] \hline
\end{tabular}}
\caption{The complexity and memory requirements during the computation of 
the matrix--vector product $\chi^0(\omega)z$. There are steps in the sequence 
with $O(N^2)$ and $O(N^3)$ complexity, where $N$ is the number of atoms.
$N_{\omega}$ denotes the number of frequencies for which the polarizability
is computed, and the remaining symbols are explained in the text.}
\label{t:flops}
\end{table}

In the first stage of the algorithm, we compute an auxiliary object $\alpha
_{\nu }^{cE}\equiv V_{\nu }^{cd}X_{d}^{E}$. The vertex $V_{\nu }^{cd}$ is a
sparse object by construction. Therefore, we will spend only $%
N_{occ}n_{orb}^{2}N_{prod}$ operations, where $N_{occ}$ is the number of
occupied orbitals, $N_{prod}$ is the number of products and $n_{orb}$ is the
number of orbitals that belong to a pair of atoms. The
auxiliary object $\alpha_{\nu }^{cE}$ is sparse and
frequency independent.
Therefore we store $\alpha_{\nu }^{cE}$ and reuse it in the
5-th step of the matrix--vector product. The product 
$\beta ^{cE}\equiv \alpha_{\nu }^{cE}v^{\nu }$ will cost $N_{occ}n_{orb}N_{prod}$
operations because each product index $\nu $
``communicates'' with the atomic orbital
index $c$ in one or two atoms. The matrix $\beta^{cE}$ will be full,
therefore the product $a^{FE}\equiv X_{c}^{F}\beta ^{cE}$ will cost $%
N_{occ}N_{virt}N_{orb}$  i.e. this step has $O(N^{3})$ complexity with $N_{virt}$
the number of unoccupied (virtual) states. The next
step $\gamma _{b}^{E}\equiv \sum_{F}X_{b}^{F}a^{FE}$ also has $O(N^{3})$
complexity with the same operation count. Finally, the sum $n_{\mu }\equiv
\sum_{E}X_{a}^{E}V_{\mu }^{ab}\gamma _{b}^{E}$ takes only $O(N^{2})$
operations because the vertex $V_{\mu }^{ab}$ is sparse. The sequence
involves $O(N^{3})$ operations, but due to prefactors, the
run time is dominated by the $O(N^{2})$ part of the sequence for molecules
of up to a hundred atoms.

\section{Calculation of the interaction kernels}
\label{s:kernels}

In the adiabatic local-density approximation (ALDA), the interaction kernel
$$f_{\mathrm{H}}+f_{\mathrm{xc}}=\frac{\delta V_{\mathrm{H}}}{\delta n}+\frac{\delta V_{\mathrm{xc}}}{\delta n}$$
is a frequency-independent matrix. While the representation of the Hartree kernel is 
straightforward in our basis
\begin{equation}
f_{\mathrm{H}}^{\mu \nu }=\iint d^{3}rd^{3}r^{\prime }\,F^{\mu }(\bm{r})%
\frac{1}{|\bm{r}-\bm{r}^{\prime }|}F^{\nu }(\bm{r}^{\prime }),
\label{hartreekerneldef}
\end{equation}%
the exchange--correlation kernel is not known explicitly and must be approximated.
The locality of the LDA potential $V_{\mathrm{xc}}(\bm{r})=V_{\mathrm{xc}}(n(\bm{r}))$ leads
to a simple expression for the exchange--correlation kernel 
\begin{equation}
f_{\mathrm{xc}}^{\mu \nu }=\int d^{3}r\,F^{\mu }(\bm{r})f_{\mathrm{xc}}(\bm{r}%
)F^{\nu }(\bm{r}),  \label{xc-kernel-def}
\end{equation}%
where $f_{\mathrm{xc}}(\bm{r})$ is a (non linear) function of the density $n(%
\bm{r})$. In this section, we describe how the Hartree and exchange-correlation kernels are constructed.

\subsection{The Hartree kernel}
\label{ss:coul-kernel} 

The basis functions $F^{\mu }(\bm{r})$ that appear
in the interaction kernels (\ref{hartreekerneldef}) and (\ref%
{xc-kernel-def}) are built separately for each pair of atoms. They are
either \textit{local} or \textit{bilocal} depending on
whether the atoms of the pair coincide or not. While the local products
are spherically symmetric functions, the bilocal products possess only
axial symmetry. Because of their axial symmetry, the bilocal products 
have a particularly simple representation in a \textit{rotated} coordinate
system
\begin{equation}
F^{\mu }(\bm{r})=\sum_{j=0}^{j_{\mathrm{cutoff}}}
F_{j}^{\mu }(r^{\prime })\text{Y}_{jm}(\bm{R}
\bm{r}'),\ \bm{r}'=\bm{r}-\bm{c},
\label{biloc-funct}
\end{equation}%
where the rotation $\bm{R}$ and the center $\bm{c}$ depend on the atom pair.
In the rotated frame, the $Z$-axis coincides with the line connecting
the atoms in the pair.
We use radial products $F_{j}^{\mu }(r)$ that are given on a logarithmic grid.
The local products have a simpler, LCAO-like representation 
\begin{equation}
F^{\mu }(\bm{r})=F^{\mu }(|\bm{r}-\bm{c}|)\textrm{Y}_{jm}(\bm{r}-\bm{c}),
\label{loc-prod}
\end{equation}%
where the centers $\bm{c}$ coincide with the center of the atom.

Using the algebra of angular momentum, we can get rid of the rotations $\bm{R}$
in the bilocal basis functions
(\ref{biloc-funct}) and transform
the kernel (\ref{hartreekerneldef}) into a sum over two center Coulomb integrals 
$\textrm{Cb}(1,2)$ 
\begin{equation}
\textrm{Cb}(1,2)=\iint d^{3}r_{1}d^{3}r_{2}\ \mathrm{g}_{l_{1}m_{1}}(%
\bm{r}_{1}-\bm{c}_{1})|\bm{r}_{1}-\bm{r}_{2}|^{-1}\mathrm{g}_{l_{2}m_{2}}(%
\bm{r}_{2}-\bm{c}_{2}).  \label{cc-coord}
\end{equation}%
The elementary functions $\mathrm{g}_{lm}(\bm{r})=\mathrm{g}_{l}(r) \mathrm{Y}_{lm}(\bm{r})$
have a radial-angular
decomposition similar to local products (\ref{loc-prod}). The Coulomb interaction
(\ref{cc-coord}) becomes local in momentum space 
\begin{equation}
\textrm{Cb}(1,2)=\int d^{3}p\ \mathrm{g}_{l_{1}m_{1}}(\bm{p})p^{-2}%
\mathrm{e}^{\mathrm{i}\bm{p}(\bm{c}_{1}-\bm{c}_{2})}\mathrm{g}_{l_{2}m_{2}}(%
\bm{p}),  \label{cc-mom}
\end{equation}%
where the Fourier image of an orbital $\mathrm{g}_{lm}(\bm{p})$ conserves  its spherical symmetry
\begin{equation}
\mathrm{g}_{lm}(\bm{p}) =\mathrm{i}\,\mathrm{g}_{l}(p)\mathrm{Y}_{lm}(%
\bm{p}).\   \label{elem-mom}
\end{equation}%
The radial part $\mathrm{g}_{l}(p)$
is given by the Hankel transform of the original radial orbital in coordinate space 
\begin{equation}
\mathrm{g}_{l}(p) =\sqrt{\frac{2}{\pi }}\int_{0}^{\infty }\mathrm{g}%
_{l}(r)\ \mathrm{j}_{l}(pr)\,r^{2}dr.
\label{hankel}
\end{equation}
Inserting the expression (\ref{elem-mom}) 
into equation (\ref{cc-mom}),
expanding the plane wave $\mathrm{e}^{\mathrm{i}\bm{p}(\bm{c}_{1}-\bm{c}%
_{2})}$ in spherical harmonics and using the algebra of angular momentum, we
can reduce the integration in momentum space to a one-dimensional integral 
\begin{equation}
I_{l_{1},l_{2},l}(R)=\int_{0}^{\infty }\mathrm{g}_{l_{1}}(p)\ \mathrm{j}%
_{l}(pR)\ \mathrm{g}_{l_{2}}(p)dp,  \label{rad-int-mom}
\end{equation}%
where $R\equiv |\bm{c}_{1}-\bm{c}_{2}|$. 

We distinguish two cases in treating this integral according to whether
the orbitals $\mathrm{g}_{lm}(\bm{r})$ are overlapping or
not. For overlapping orbitals, we compute the
integral (\ref{rad-int-mom}) numerically using Talman's fast Hankel transform \cite{Talman:1983:2009}. 
For non overlapping orbitals, one can compute the integral  (\ref{rad-int-mom}) exactly using
a multipole expansion. To derive this expansion, we replace
the functions $\mathrm{g}_{l}(p)$ in the equation (\ref{rad-int-mom}) by
their Hankel transforms (\ref{hankel}) 
\begin{equation}
I_{l_{1},l_{2},l}(R)=\frac{2}{\pi }\int_{0}^{\infty }dr_{1}\,\mathrm{g}%
_{l_{1}}(r_{1})r_{1}^{2}\int_{0}^{\infty }dr_{2}\,\mathrm{g}%
_{l_{2}}(r_{2})r_{2}^{2}\int_{0}^{\infty }\mathrm{j}_{l_{1}}(r_{1})\,\mathrm{%
j}_{l}(pR)\,\mathrm{j}_{l_{2}}(r_{2})\,dp.  \label{rad-3int-mom}
\end{equation}%
The integral over three spherical Bessel functions $%
I_{l_{1},l_{2},l}(r_{1},r_{2},R)\equiv \int_{0}^{\infty }\mathrm{j}%
_{l_{1}}(r_{1})\mathrm{j}_{l}(pR)\mathrm{j}_{l_{2}}(r_{2})\,dp$ reduces to a 
simple separable expression  \cite{Gradsteyn} provided two conditions are
satisfied, $R>r_{1}+r_{2}$ (the basis functions do not overlap), and $0\leq
l\leq l_{1}+l_{2}$ (the triangle relation for angular momentum). Under these
conditions, we obtain
\begin{equation}
I_{l_{1},l_{2},l}(r_{1},r_{2},R)=\delta _{l,l_{1}+l_{2}}\frac{\pi ^{3/2}}{8}%
\frac{r_{1}^{l_{1}}r_{2}^{l_{2}}}{R^{l+1}}\frac{\Gamma (l+1/2)}{\Gamma
(l_{1}+3/2)\Gamma (l_{2}+3/2)}.
\end{equation}%
Inserting this result into equation (\ref{rad-3int-mom}), we obtain
an expression for the Coulomb interaction in terms of moments in closed form 
\begin{equation}
\rho _{l}\equiv \int_{0}^{\infty }r^{2}dr\,\mathrm{g}_{l}(r)r^{l}.
\label{moment}
\end{equation}%
The moments (\ref{moment}) can be computed and stored at the beginning of
the calculation. Therefore, the calculation will consist of summing the
angular--momentum coefficients and this is clearly much faster than a direct
numerical integration in equation (\ref{rad-int-mom}).

The complexity of the near-field interactions will be proportional to the number of
atoms $N$. The calculation of the multipoles (\ref{moment}) for the
far-field interaction requires of $O(N)$ mathematical operations.
The remaining part of the far-field interactions (Wigner rotations) scales
as $N^{2}$ with the number of atoms. 

\subsection{The exchange--correlation kernel}
\label{ss:xc-kernel}

Unlike the Hartree kernel, the exchange-correlation kernel (\ref{xc-kernel-def})
is local and we therefore compute it directly in coordinate space by numerical quadrature. 
The support of the dominant products $F^{\mu }(\bm{r})$, $F^{\nu }(\bm{r})$
in the integrand of equation (\ref{xc-kernel-def}) have, in general, the shape of a lens.
Therefore, the support of the integrand will generally be an intersection
of two lenses. 

\begin{figure}[tbp]
\centerline{\includegraphics[width=4.2cm, angle=0,clip]{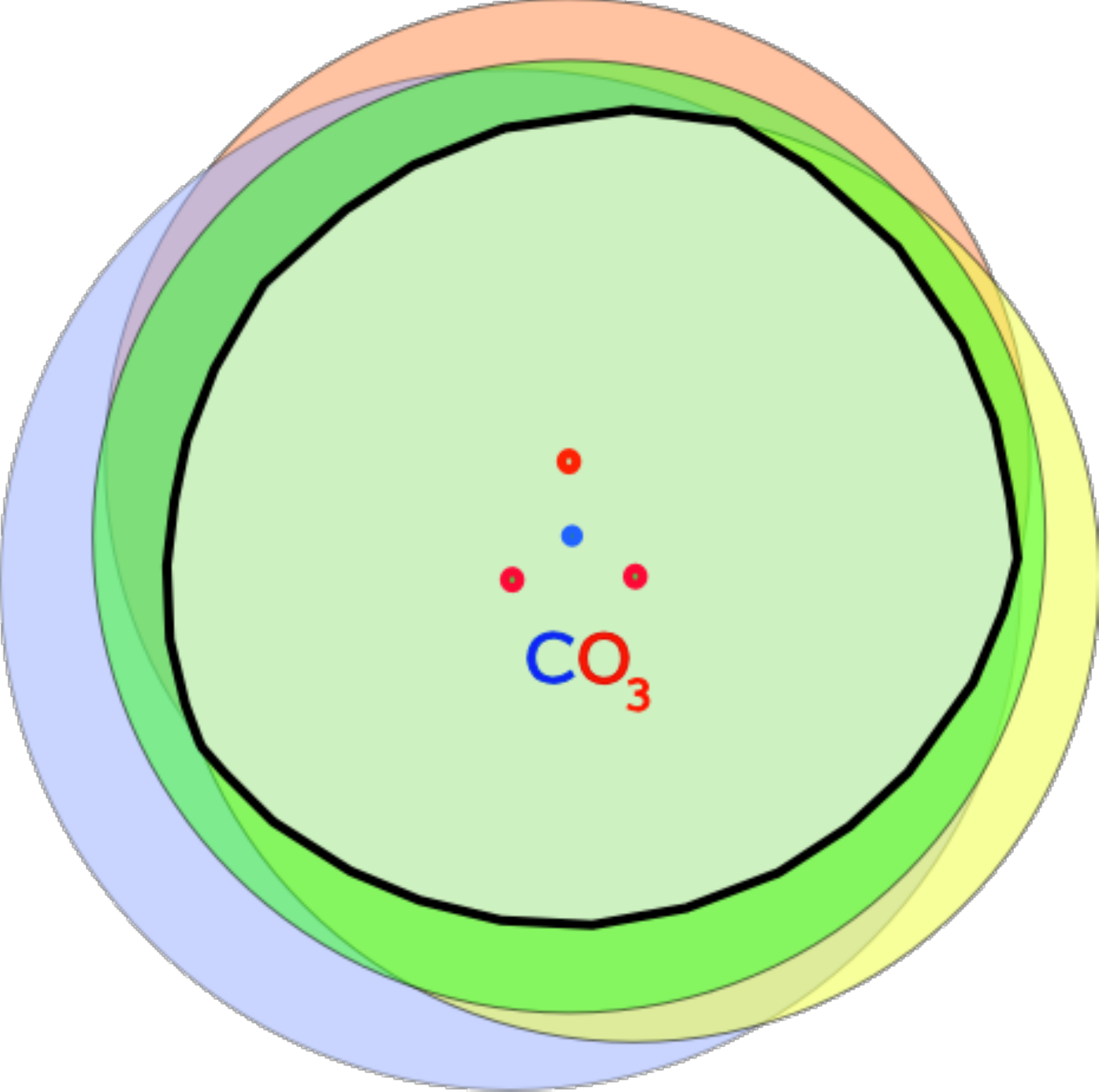}}
\caption{
The spatial support of the integrand of the exchange correlation term (\ref{xc-kernel-def})
depends on the support of its underlying atomic orbitals and on the geometry of
the quadruplet of atoms under consideration. For neighboring atoms, the support of the orbitals
is several times larger than their inter atomic distances and this results in a nearly 
spherical support of the integrand. The figure illustrates the case of (hypothetical) CO$_3$}
\label{f:xc-geom}
\end{figure}

We found, however, that integration in spherical coordinates
gives sufficiently accurate and quickly convergent results. This is because
the important matrix elements involve neighboring dominant products
and the support of the dominant products is large compared to the distance
between their centers. Therefore, the integrands of the important matrix
elements (\ref{xc-kernel-def}) have nearly spherical support. This situation
is illustrated by the cartoon in figure \ref{f:xc-geom}.

For each pair of dominant products, we use spherical coordinates that are centered
at the midpoint between them. We use Gauss-Legendre quadrature for integrating along
the radial coordinate and Lebedev quadrature \cite{Lebedev-theory-and-program} for
integrating over solid angle
\begin{equation}
F_{\mathrm{xc}} = \sum_{i=1}^{N_r} G_i \sum_{j=1}^{N_{\Omega}} L_j \, f_{\mathrm{xc}}(r_i, \Omega_{j}).
\end{equation}
Here $G_i$ and $r_i$ are weights and knots of Gauss-Legendre quadrature, and $L_j$ and 
$\Omega_{j}$ are weights and knots of Lebedev quadrature. The number of points $N_r\times N_{\Omega}$
can be kept reasonably small ($24 \times 170$ by default).

The most time consuming part of the exchange-correlation kernel is the 
electronic density. We found that calculating the density using the density matrix
\begin{equation}
n(\bm{r}) = \sum_{ab} f^a(\bm{r}) D_{ab} f^b(\bm{r}), \text{ where } D_{ab} = \sum_{E<0} X^E_a X^E_b
\end{equation}
provides a linear scaling of the run time of $f_{\mathrm{xc}}$ with the number of atoms. However, a calculation of the
density via molecular orbitals
\begin{equation}
n(\bm{r}) = \sum_{E<0} \psi_E(\bm{r}) \psi_E(\bm{r}), \text{ where } \psi_E(\bm{r}) = \sum_{a} X^E_a f^a(\bm{r})
\end{equation}
is faster in many cases, although the run time scales quadratically with the number of atoms.
In order to optimize the run time, rather than insist on linear scaling, we choose the calculational
approach automatically depending on the geometry of the molecule. For instance, in the case of a long polythiophene 
with 13 chains (see subsection \ref{ss:complexity}) we use the $O(N)$ method, while in the other examples
it is better to use the $O(N^2)$ approach.

\section{Parallelization of the algorithm}
\label{s:parallelization}

The overall structure of our algorithm is given in figure~\ref{a:main}.
First, the basis of dominant products is built, then the  interaction kernels are computed,
and finally both are used in the iterative procedure to compute the dynamical polarizability.

\begin{algorithm}[htbp]
\begin{center}
\fcolorbox{green}{white}{
  \parbox{8cm}{
    \Indm\textbf{Construction of the dominant products} \\
      \Indp
         \For{atom species}{Build local dominant products}
         {\scriptsize!\$OMP PARALLEL DO}\\
         \For{atom pairs $(a,b)$}{Build bilocal dominant products} 
  }
}
\fcolorbox{red}{white}{
  \parbox{8cm}{
    \Indm\textbf{Computation of the interaction kernels $f_{\mathrm{H}}$ and $f_{\mathrm{xc}}$}\\
    \Indp
      {\scriptsize!\$OMP PARALLEL DO} \\
      \For{each couple of atom pairs $(a,b;c,d)$}{
            Build block $(a,b;c,d)$ of kernels.}
  }
}
\fcolorbox{blue}{white}{
  \parbox{8cm}{
  \Indm\textbf{Computation of the dynamical polarizability}\\
     \Indp
         {\scriptsize!\$OMP PARALLEL DO} \\
         \For{$\omega\in[\omega_{\mathrm{begin}},\omega_{\mathrm{end}}]$}{
            Solve for $|X\rangle$: $\left(1-\chi^0(\omega) f_{\mathrm{Hxc}}\right)|X\rangle = \chi^0(\omega)| d\rangle$ \\
            Compute polarizability $P(\omega) = \langle d|X\rangle$ }
  }
}
\end{center}
\caption{Skeleton of the algorithm. Its first (and computationally easiest) part is the
construction of the dominant products. The second (and computationally most demanding) part 
is the construction of the interaction kernels. The third (and  comparatively easy) part is the 
iterative calculation of the dynamical polarizabilities $P(\omega)$.}
\label{a:main}
\end{algorithm}

The individual components of the algorithm~\ref{a:main} suggest different parallelization
strategies. The dominant products are built for each atom pair independently, therefore
the corresponding code is parallelized over atom pairs. The structure of the dominant
products suggests a block wise computation of the interaction kernels. These blocks are
mutually independent, therefore we parallelize
their construction. The dynamical polarizabilities are calculated independently for each
frequency and are, therefore, parallelized over frequencies.
Below, we go through the details of the algorithm and its hybrid OpenMP/MPI parallelization.

\subsection{Multi-thread parallelization}

Modern computers are faster than previous generations of machines mainly
due to their parallel design as in the case of general-purpose multi-core processors.
For specially written programs, such a design allows to run
several tasks or ``threads'' simultaneously. Fortunately, it is easy to write 
a multi-threaded program using the current application programming interface OpenMP. 
Moreover, in OpenMP, data exchange between threads uses the common memory and it is, therefore,
faster than on distributed-memory machines. 
For these reasons, our main emphasis here is on multi-threaded (or shared-memory) parallelization.
We use the OpenMP standard \cite{OpenMP} that allows for an efficient parallelization
of all three sections of the algorithm~\ref{a:main}.

\subsubsection{Building the basis of dominant products}
\label{sss:domi-prod}

The construction of local dominant products involves only the atomic species that
occur in a molecule. Therefore it is computationally cheap and any parallelization
would only slow down their construction.

For bilocal dominant products, the situation is different. The construction of bilocal 
dominant products is done for all atom pairs, the orbitals of which overlap. Because these pairs
are independent of each other, we parallelize the loop over pairs with OpenMP directives. 

The dominant products $F^{\mu}(\bm{r})$ and vertices $V_{\mu}^{ab}$ are stored
in suitably chosen data structures to allow for effective use of memory. 
Due to the locality of
our construction, the amount of memory spent in the storage of dominant products and vertices grows linearly 
with the number of atoms.

\subsubsection{Construction of the interaction kernels}
\label{sss:omp-kernels}

\begin{figure}[htbp]
\centerline{\includegraphics[width=7cm,angle=0,clip]{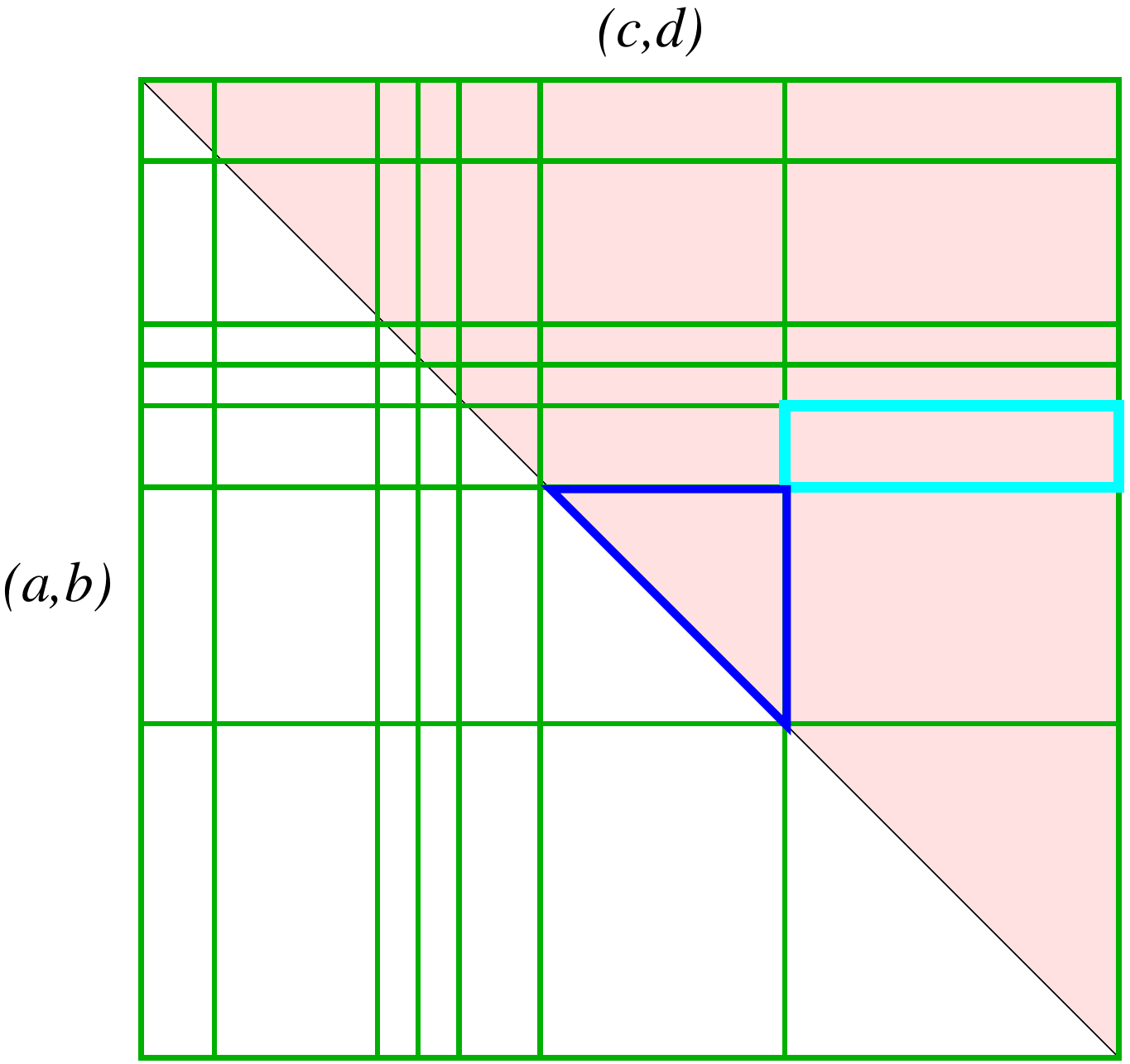}}
\caption{The block structure of the interaction kernels. The blocks are defined by two atom pairs
$(a,b)$ and $(c,d)$. The size of blocks is known from the construction of the  
dominant products. Because the entire interaction kernel $f^{\mu\nu}_{\mathrm{Hxc}}$  
is symmetric, only its upper triangular part is computed.}
\label{f:block-structure}
\end{figure}

The interaction kernels (\ref{hartreekerneldef}) and (\ref{xc-kernel-def}) refer to a pair 
of dominant products $F^{\mu}(\bm{r})$ and $F^{\nu}(\bm{r})$. Because each of the dominant products
in turn refers to a pair of atoms, the interaction kernel splits into blocks that are labeled 
by a quadruplet of atoms $(a,b;c,d)$. In practice, this is used to precompute and reuse some 
auxiliary quantities that belong to such a quadruplet.
The block structure is schematically depicted in
figure \ref{f:block-structure}. Generically, a block is rectangular, but when two pairs coincide,
the block reduces to a triangle because of reflection symmetry.

The calculation of the interaction kernels is parallelized over blocks using dynamical scheduling.
Because the computational load of each block is proportional to its area,
we minimize the waiting time at the end of the loop by a descending 
sort of the blocks according to their area.

\subsubsection{Computation of the dynamical polarizability}
\label{sss:dynamical-polariz}

According to equation (\ref{eqpolarizability}), the dynamical polarizability $P_{ik}(\omega)$
is computed independently for each frequency $\omega$. Therefore, the loop
over frequencies in the algorithm~\ref{a:main} is embarrassingly parallel.

In the parallelization over frequencies, we had to make thread safe
copies of module variables (working arrays in the GMRES solver) using the 
OpenMP directive \texttt{threadprivate}. This multiplies the memory requirement
by the number of threads, but this poses no problem, because this part of the algorithm
is not memory intensive.

The number of iterations to reach convergence of the polarizability tensor (\ref{eqpolarizability})
varies with frequency and with the Cartesian components in an irregular way.
To take this into account, we use a dynamic schedule with a single frequency and
tensor component per thread. By treating tensor components on the same level
as frequencies, we reduce the body of a loop by a factor of three provided only
diagonal components are computed.

\subsection{Hybrid MPI-thread parallelization}

Most current supercomputers are organized as clusters of multi-core nodes that are
interconnected by a high speed network. Although the number of cores grows over the years,
we still need several nodes for greater computational speed and to provide sufficient memory.

Our program has also been adapted to such distributed-memory parallel machines. 
We parallelize according to the Single Program Multiple Data (SPMD) paradigm with
the message passing interface (MPI) to speed up only the computationally 
intensive parts of the algorithm --- the construction of the bilocal products,
the calculation of the kernels $f_{\mathrm{H}}$, $f_{\mathrm{xc}}$ and the iterative procedure.
Moreover, each MPI process uses the multi-thread parallelism described above.

\subsubsection{Parallelization of the basis of dominant products}

Only bilocal dominant products must be constructed in parallel.
As described in subsection \ref{sss:domi-prod}, the construction is naturally 
parallel in terms of atom pairs. Therefore, we distribute the atom pairs prior to
computation and gather data after the computation in order to duplicate basis functions
on each MPI process.

\subsubsection{Parallelization of the interaction kernels}

As explained in subsection \ref{sss:omp-kernels}, the interaction kernel depends
on quadruplets of atoms and it would appear natural to parallelize their construction over
these quadruplets. We found it advantageous, however, to slice the interaction matrix
into vertical bands that belong to one or more atom pairs and process them on different nodes.
After the computation, the complete matrix of interactions is reconstituted on each node. 

The optimal size of each  vertical band is determined by an estimate of the work load prior to the computation. 
Since the Hartree and the exchange-correlation kernel differ in their properties (such as locality) their matrices are 
distributed differently. In both cases, however, the total work load is the sum of the work loads
of its constituent quadruplets.

In the case of the Hartree kernel (\ref{hartreekerneldef}), the work load of a block depends on its size and
on whether its atom pairs are local or bilocal. While the local products are of simple LCAO type
(\ref{loc-prod}), the bilocal products (\ref{biloc-funct}) contain additional summations.
We found the following robust estimator of the workload of a block of the Hartree kernel
\begin{equation}
\mathrm{Workload}(\mathrm{Hartree}) = \mathrm{Size\_Of\_Block}
\cdot(j_{\mathrm{cutoff}}\cdot \Theta(a\ne b)+1)
\cdot(j_{\mathrm{cutoff}}\cdot \Theta(c\ne d)+1),
\end{equation}
where $\Theta(a\ne b)$ is equal to one for bilocal atom pairs and zero otherwise,
$j_{\mathrm{cutoff}}$ is the largest angular momentum in the expansion 
(\ref{biloc-funct}). By default, its upper limit is set to 7.

In the exchange-correlation kernel (\ref{xc-kernel-def}), the domain of integration is the
intersection of two lens-like regions. 
Because there is no simple analytical expression for such a volume, we count the number 
of integration points within the support for each block and estimate the work load
as proportional to the number of points. Rather surprisingly, the run time
is independent of the dimension of the block (this is due to precomputing the values of the dominant products). 

Because the kernel $f_{\mathrm{Hxc}}$ is frequency independent, it is computed at
the beginning of the iterative procedure and duplicated on all the MPI-nodes.
We found the time for gathering the matrix small compared to the time of its computation.

\subsubsection{Parallelization of the iterative procedure}

As mentioned previously in subsection \ref{sss:dynamical-polariz},
the iterative calculation of the polarizability tensor $P_{ik}(\omega)$
is naturally parallel both in its frequency and in its Cartesian tensor indices.
Therefore, this part of the algorithm is parallelized over both
frequency and tensor indices. 
However, the workload for each composite index is difficult to predict. To achieve a balanced workload
on the average, we distribute this index cyclically over MPI nodes (``round robin distribution''). 

\section{Results}
\label{s:results}

In this section, we present different absorption spectra of large molecules to validate the method
and the parallelization approach.
We start in subsection \ref{ss:fullerene} by comparing our absorption spectra with
previous calculations by other authors and also with experimental data.
Then, in subsection \ref{ss:complexity}, we present the scaling of the run time with the number of atoms.
In subsection \ref{ss:parallel-results},
we examine the efficiency of the hybrid parallelization for a variety of
molecules run on different machines. Finally, in subsection \ref{ss:fullerene-versus-pcbm}, we 
present a comparison of absorption spectra of fullerene C$_{60}$ with its derivative PCBM
that is often used in organic solar cells.

\subsection{Fullerene C$_{60}$}
\label{ss:fullerene}

We already tested the basis of dominant products in our previous works
\cite{DF:2008,DF:2009,PK-DF-OC:2010}, where the absorption
spectra of methane, benzene, indigo blue, and buckminsterfullerene C$_{60}$
were studied. The new element in the present paper is the use of an iterative method without 
constructing the full Kohn-Sham response response function. We now test this method on buckminster
fullerene and its functionalized derivative PCBM.

Buckminster fullerene C$_{60}$ is of considerable interest in materials
science. Among other applications, it is used as an electron acceptor in organic solar
cells \cite{Brabec:2001}. Here we compare our results with spectra from the Quantum Espresso
package and with experiment. 

\begin{figure}[tbp]
\centerline{\includegraphics[width=7cm,viewport=80 60 400 300, angle=0,
clip]{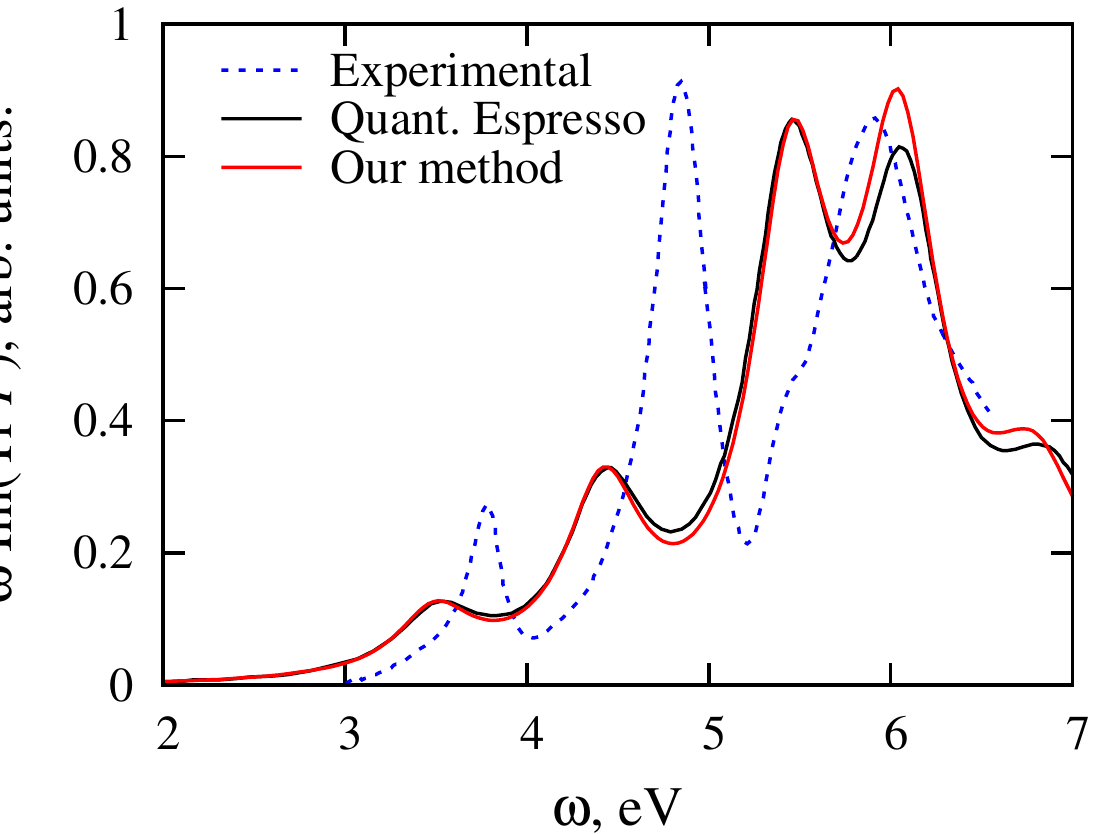}}
\caption{Comparison of the low-frequency absorption spectra of C$_{60}$
fullerene. We see a good agreement between the two theoretical predictions, while
the experimental curve is shifted by $0.2\ldots 0.3$ eV.}
\label{f:espresso}
\end{figure}

The absorption spectrum of fullerene C$_{60}$ is shown in
figure \ref{f:espresso} where we compare our result with
the calculation by Rocca \textit{et al.}
\cite{Rocca-etal:2008} and with experiment \cite{Bauernschmit:1998}.
One can see a good agreement between the theoretical spectra, while both
theoretical predictions deviate from the experimental data by $0.2\ldots 0.3$
eV. The shift in the spectrum might be due to the solvent in the experimental
setup or due to the inadequacy of the simplest LDA functional for this large
molecule. We used DFT data from the SIESTA package \cite{siesta}, where
the pseudo-potentials of the Troullier-Martin type and the LDA
functional by Perdew and Zunger \cite{Perdew-Zunger:1981} are applied.
A double-zeta polarized basis set has been used and the broadening of
levels has been set to $0.019$ Ry. Our program spent a total of about
62.5 minutes on this calculation, of which 2098 seconds were spent 
on the Coulomb kernel, 1085 seconds on the exchange-correlation kernel,
and 500 seconds in the iterative procedure, i.~e. approximately 2.27 second
per frequency. The convergence parameter for the polarizability was set to 1\%.
With this convergence parameter, the dimension of the Krylov space was varying
from 7 to 12 with an average of 8, while the dimension of the dominant product
space was 8700.
In this test, we used one thread on a machine with two CPU Intel quad core
Nehalem @ 2.93GHz; 8MB cache; 48GB DDR3 RAM,
and consuming no more than 2.3\% of RAM (1.2\% during the iterative procedure).

\subsection{Polythiophene chains (complexity of the method)}
\label{ss:complexity}

In sections \ref{s:iter} and \ref{s:kernels}, we discussed the complexity 
scaling of different parts of the algorithm theoretically. 
In this subsection, we measure the dependence of run time on the number
of atoms $N$ in polythiophene chains of different lengths. We shall see that the run time scaling
follows the theoretical predictions for the complexity.

\begin{figure}[tbp]
\centerline{
\includegraphics[width=7cm,viewport=50 60 410 300, angle=0,clip]{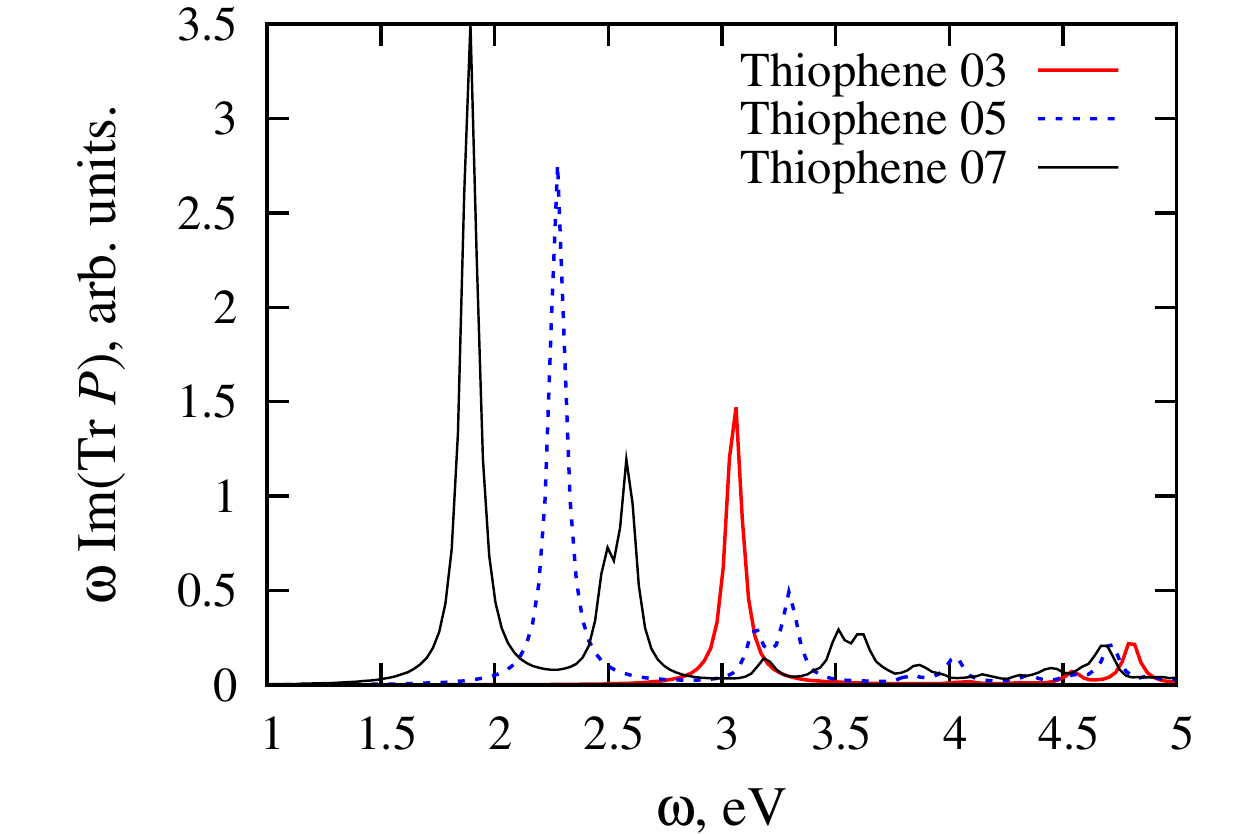}
\hspace{-0.5cm}
\includegraphics[width=7cm,viewport=50 60 410 300, angle=0, clip]{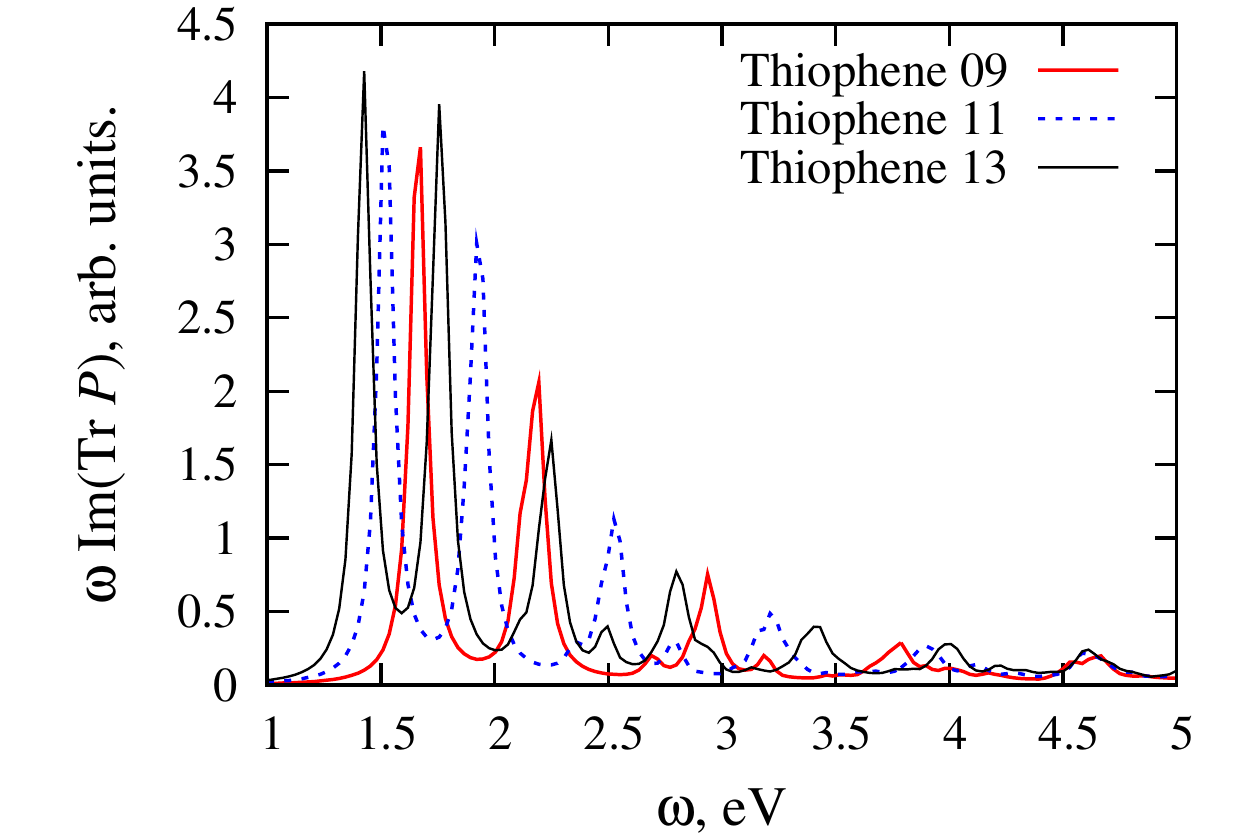}
}
\caption{Comparison of low-frequency spectra for several polythiophene chains.}
\label{f:thiophene-spectra}
\end{figure}

Sulfur containing molecules are widely use in organic electronics \cite%
{Brabec:2001,Fushun-etal:2009,Gao-etal:2010}. In this work, we study
pure polythiophene chains of 3 to 13 repeating units. The geometry of the
longest polythiophene we considered is shown in figure \ref%
{f:thiophene-13-geom}. Our calculations suggest (ignoring the excitonic character of these molecules)
that the HOMO--LUMO energy difference decreases, while the
absorption increases, with chain length. The calculated absorption spectra are collected
in figure \ref{f:thiophene-spectra}.

We now use the calculations on polythiophene spectra in order to 
study the run time scaling with the number of atoms of different parts of our algorithm. Their scaling behavior 
will be described in terms of approximate scaling exponents. The run times for a few chains are collected
in table \ref{t:runtime} for a machine of four CPU AMD Dual-Core Opteron @ 2.6GHz;
8MB cache; 32GB DDR2 RAM, and running sequentially.

The application of the non interacting response to a vector consists of $N^2$-
and $N^3$-parts (see subsection \ref{ss:matrix-vector}). The total time for
the product $\chi^0\,z$ is collected in the third column of table \ref{t:runtime},
while the run time of the $N^3$-part is collected in the fourth column. Using the 
run times, one can compute exponents $x$ and $x_3$ for their corresponding scaling laws 
$N^x$ and $N^{x_3}$. The exponents $x$ and $x_3$ vary in the range of  $x=2.31\ldots 2.36$
and $x_{3}=2.49\ldots 2.53$, respectively. The run time of the
Hartree kernel (fifth column)
shows a scaling exponent in the range $x_{\mathrm{H}}=2.05\ldots 2.06$, while the
run time in the exchange--correlation kernel (sixth column) scales almost
linearly $x_{\mathrm{xc}}=1.04\ldots 1.12$. Therefore, the measured exponents are
close to the predicted exponents $x=3$, $x_{3}=3$,
$x_{\mathrm{H}}=2$, and $x_{\mathrm{xc}}=1$.

\begin{figure}[tbp]
\centerline{\includegraphics[width=15cm,viewport=50 360 1500 560, angle=0,
clip]{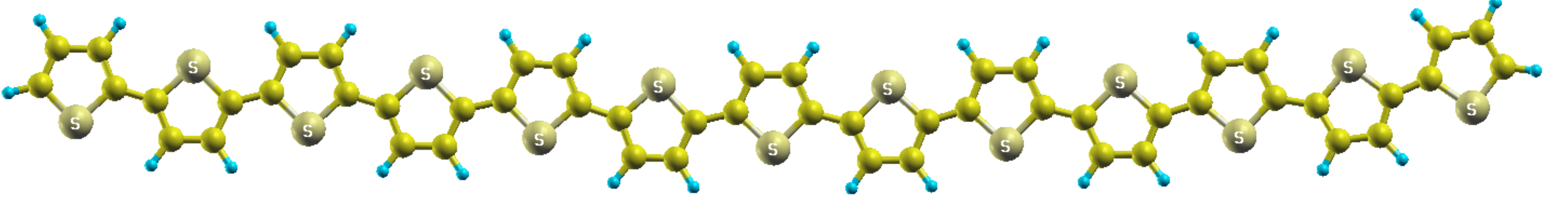}}
\caption{The geometry of the longest polythiophene chain we considered.}
\label{f:thiophene-13-geom}
\end{figure}

\begin{table}[htb]
\begin{center}
\begin{tabular}{|c|c||c|c|c|c|}
\hline
$N$& Units & $\chi^0 z$, s 
                      & $N^3$ part in $\chi^0 z$, s 
                                 & $f_{\mathrm{H}}$,s  
                                        & $f_{\mathrm{xc}}$, s \\ \hline
23 & 3     & 1.51E-02 & 8.20E-03 & 157  & 173  \\
37 & 5     & 4.42E-02 & 2.52E-02 & 417  & 300  \\
51 & 7     & 9.29E-02 & 5.58E-02 & 807  & 428  \\
65 & 9     & 0.166    & 0.104    & 1324 & 561  \\
79 & 11    & 0.260    & 0.169    & 1977 & 694  \\
93 & 13    & 0.382    & 0.255    & 2767 & 822  \\
\hline
\end{tabular}
\end{center}
\caption{Run time in different parts of the algorithm as a function of the number
of atoms $N$ in the polythiophene chain. The third column gives the total
time for the matrix--vector product $(\chi^0 z)$, while an $N^3$-part
that arises in the construction of $\chi^0 z$ is given in the fourth column. The fifth and sixth columns
display the run times of the interaction kernels $f_{\mathrm{H}}$
and $f_{\mathrm{xc}}$, respectively.}
\label{t:runtime}
\end{table}

The calculation of the Hartree kernel $f_{\mathrm{H}}$ via multipoles, as
explained in subsection \ref{ss:coul-kernel} improves the
run time in the case of large molecules, but could not improve the run time
scaling of the Coulomb interaction. In fact, the Hartree kernel $f_{\mathrm{H}}$ is a non-local
quantity and the rotations involved in the bilocal dominant products contribute a
substantial part to the run time. 

The scaling of the run time for the entire
calculation of molecular spectra will also vary with the parameters of the calculation.
Obviously, for a small number of frequencies the scaling of total run time will
be determined by the Hartree and exchange-correlation kernels. However,
if the number of frequencies is large, then the application of the non
interacting response $\chi^{0}$ on a vector will dominate the run time.

\subsection{Quality of the parallelization}
\label{ss:parallel-results}

Our parallel Fortran 90 code is adapted to 
current parallel architectures (see section \ref{s:parallelization}).
In this subsection, we evaluate the quality of the hybrid parallelization
by testing our approach on three machines of different architecture. Two machines belong to 
shared-memory multi-core architectures with non uniform memory access,
while the third machine is a cluster with 50 multi-core nodes interconnected by
an Infiniband network.

The program was compiled using Intel's Fortran compiler and linked against
Intel's Math Kernel Library for BLAS, LAPACK and Fast Fourier Transform libraries.

\subsubsection{Multi-thread parallelization}
\label{sss:smp}

\begin{figure}[htbp]
\centerline{\includegraphics[width=6cm, angle=0, clip]{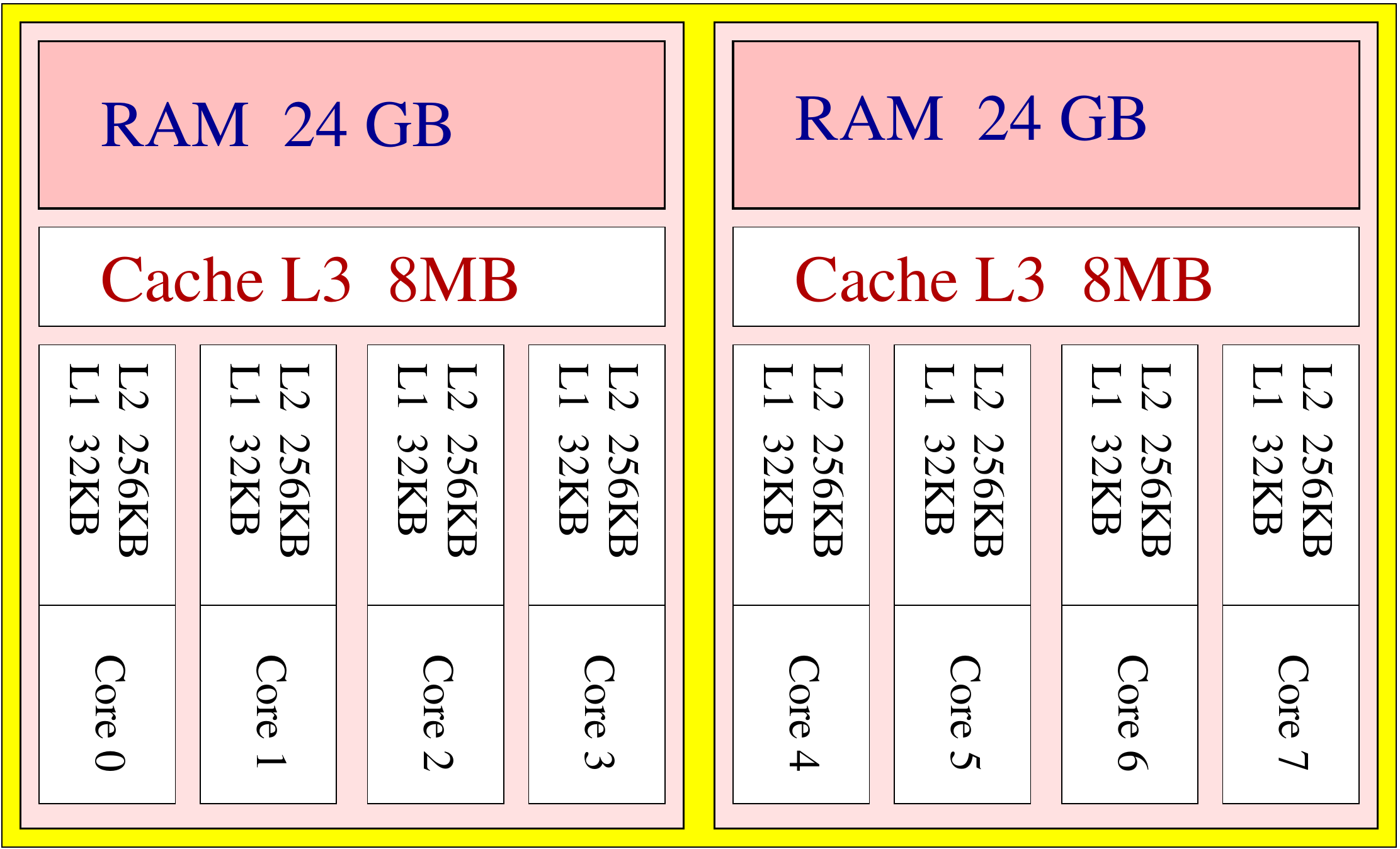}}
\caption{Memory/cache/core structure of Nehalem Intel Xeon X5550 machine. Two
quad core nodes have a fast access to one of the memory banks, while the
inter node communication is slower.}
\label{f:nehalem-machine}
\end{figure}

\begin{figure}[htbp]
\centerline{\includegraphics[width=6cm, angle=0, clip]{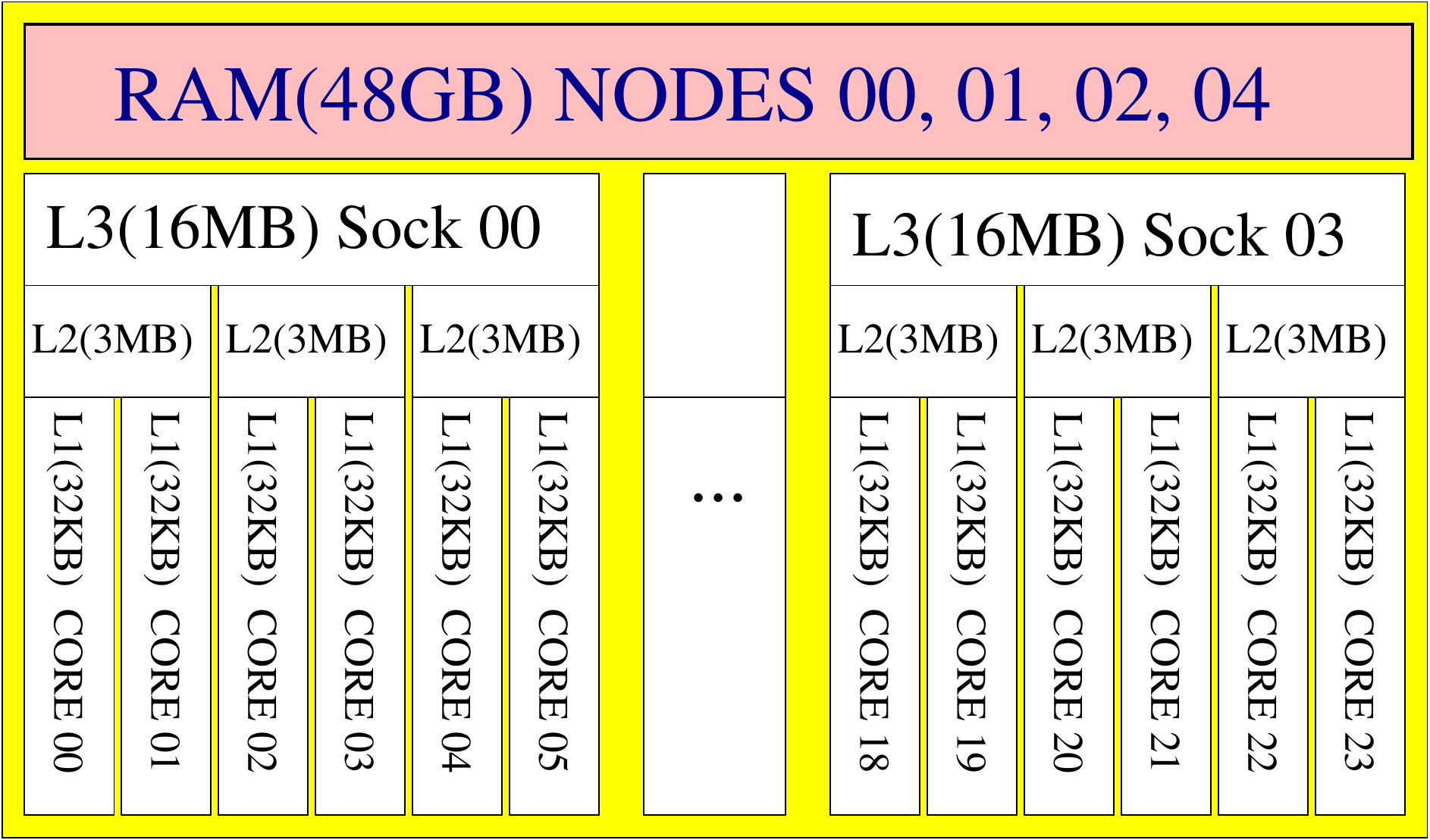}}
\caption{Memory/cache/core structure of Xeon-96 machine. 
48 dual core Xeon CPUs are connected to four memory banks. Although
every core can address the whole memory space, the inter node
communication is slower.}
\label{f:xeon-96-machine}
\end{figure}

We consider parallelization on shared-memory machines as more
important than on distributed-memory machines. 
Therefore, the speed up of our code is first tested on two shared-memory machines with 
Non-Uniform Memory Architecture (NUMA). The first machine has two quad core Nehalem 5500
CPUs (see figure \ref{f:nehalem-machine}),
while the second machine has 48 dual core Xeon CPUs (see figure \ref{f:xeon-96-machine}).

\begin{figure}[htb]
\centerline{
\includegraphics[width=7cm,viewport=50 60 410 300, angle=0,clip]{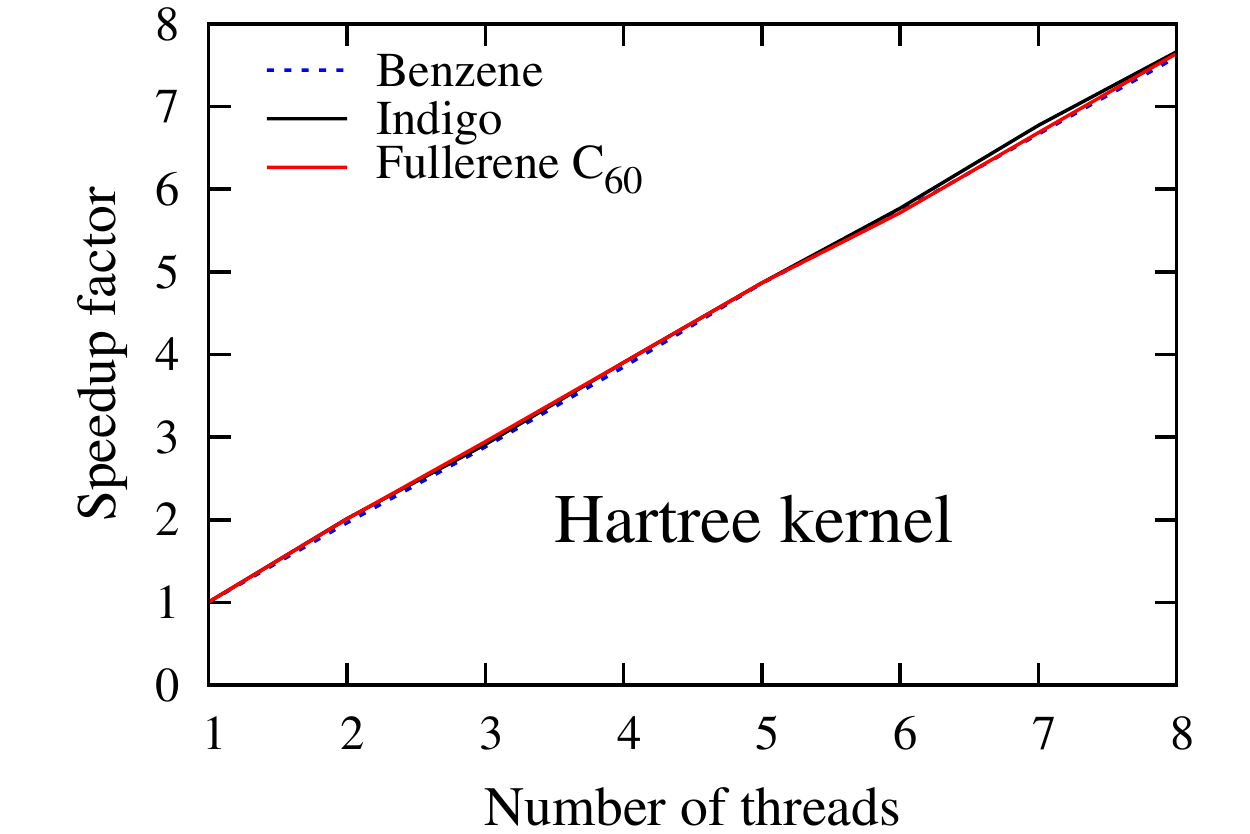}
\hspace{-0.5cm}
\includegraphics[width=7cm,viewport=50 60 410 300, angle=0, clip]{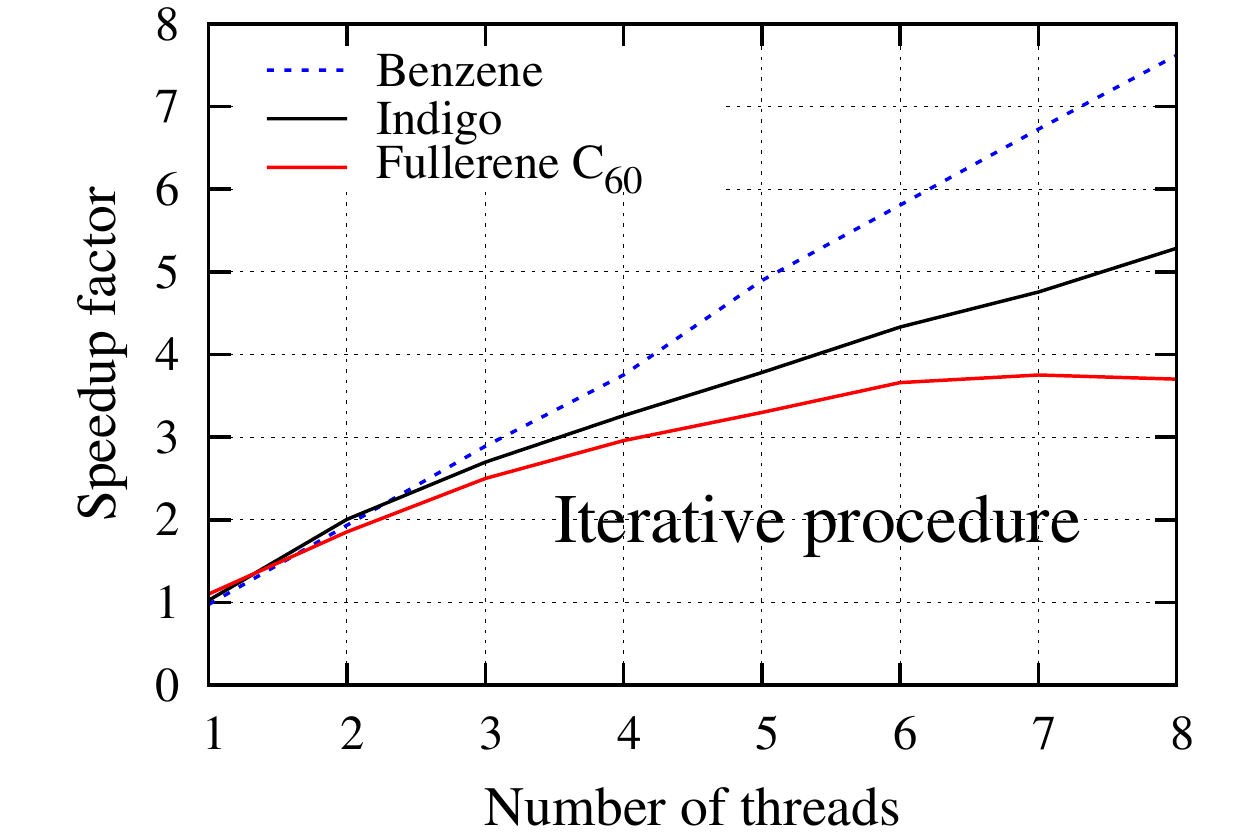}}
\caption{The left panel shows the speedup factor in the Hartree kernel for molecules of different size.
The speedup in the exchange-correlation kernel and in the generation of dominant products is similar
to the left panel. The right panel shows the speedup factor in the iterative procedure. We observe 
that the speed up decreases with the size of the molecule.}
\label{f:speedup-omp-molecule-size}
\end{figure}

The speedup on the Nehalem machine is shown in figure \ref{f:speedup-omp-molecule-size}
for three molecules of different size: benzene, indigo and fullerene. The speedup in
both kernels and in the generation of bilocal products is good for all molecules, therefore
we plot only the speedup for the Hartree kernel on the left panel. By contrast, the iterative procedure
exhibits a lower speedup (as shown on the right panel in figure \ref{f:speedup-omp-molecule-size}).
This is due to the high memory-bandwidth requirement of the iterative procedure.
The high memory-bandwidth is clearly revealed in two distinct ways of 
using the same number of cores in a test calculation, see subsection \ref{sss:mpi-results}.

\begin{figure}[htb]
\centerline{
\includegraphics[width=7cm,viewport=50 60 410 300, angle=0,clip]{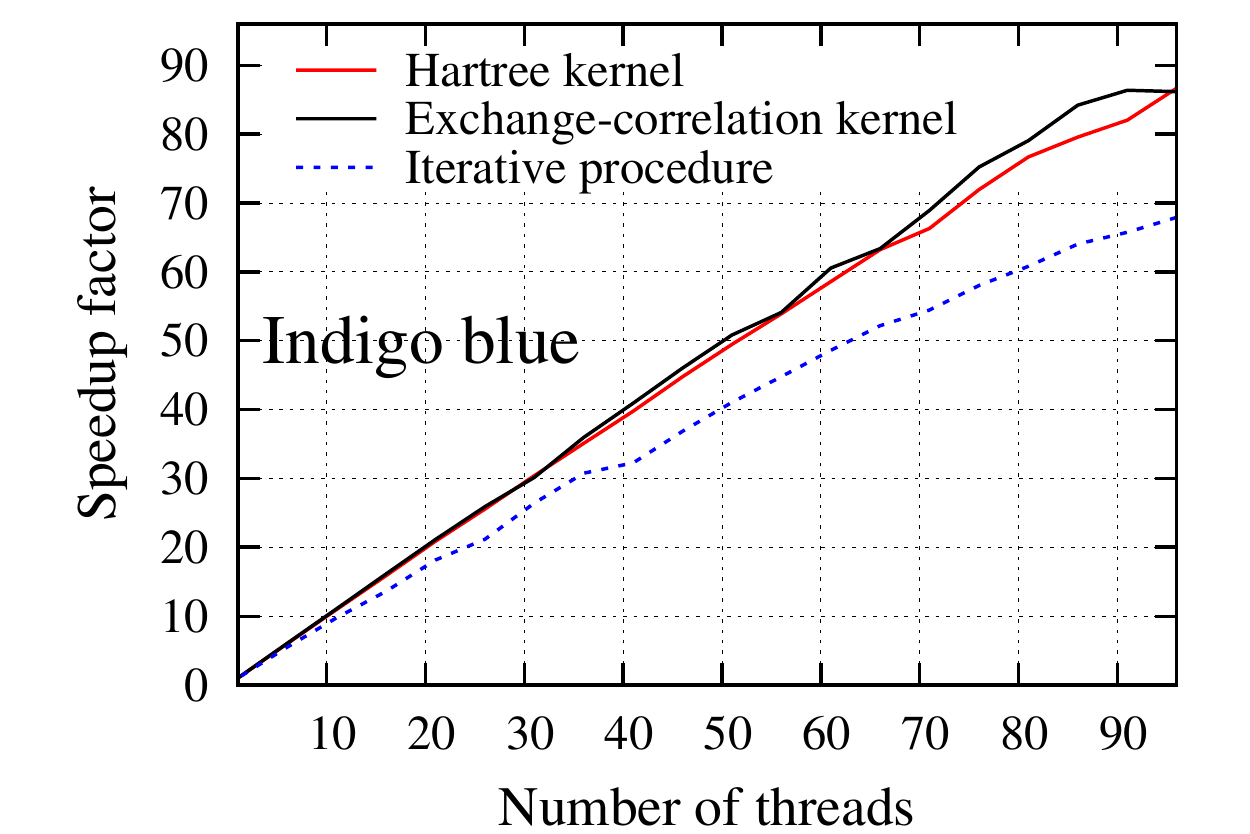}
\hspace{-0.5cm}
\includegraphics[width=7cm,viewport=50 60 410 300, angle=0,clip]{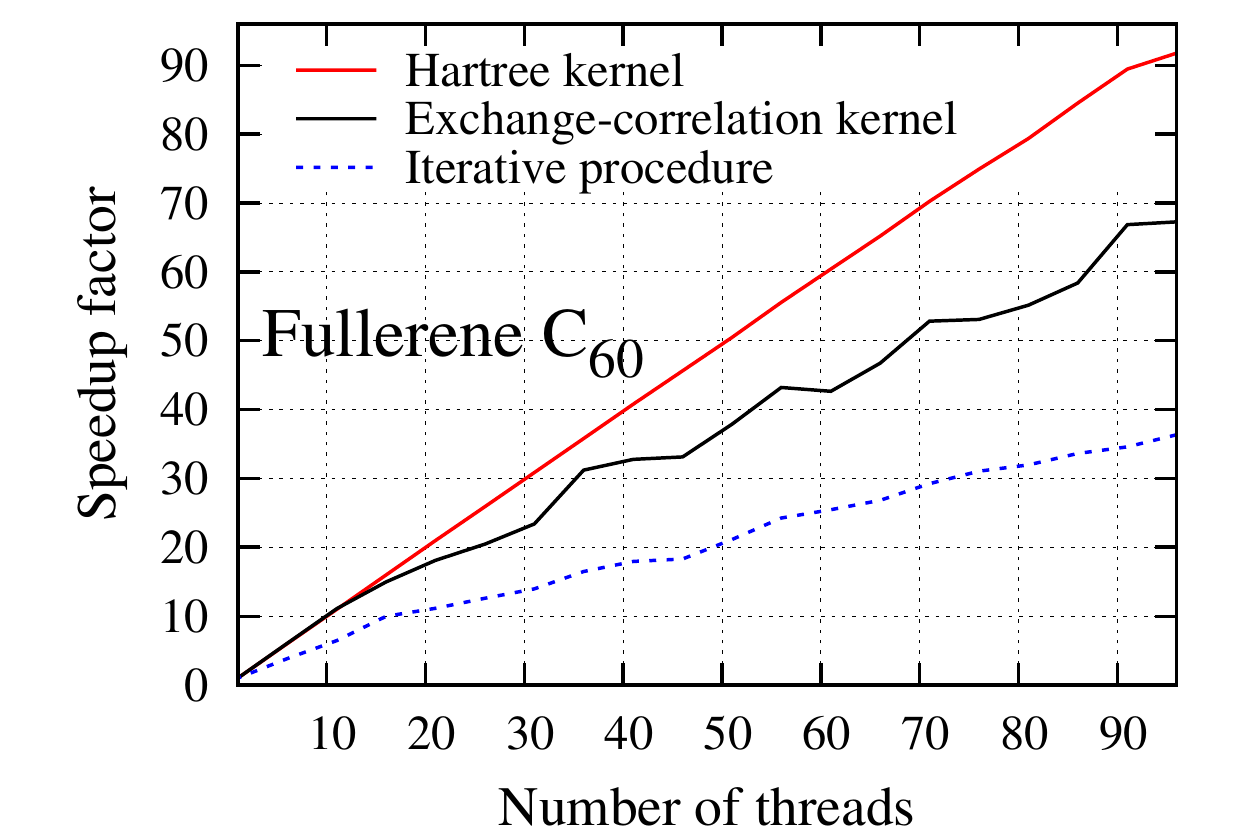}
}
\caption{Speedup on a heavily parallel Xeon-96 machine. The results are satisfactory for
the interaction kernels, while the memory-bandwidth requirements of the iterative procedure
hamper the parallel performance in the case of the larger fullerene C$_{60}$ molecule.}
\label{f:speedup-omp-bertha-molecule-size}
\end{figure}

In spite of the loss of speed up for larger molecules on the Nehalem machine, we tested
our implementation also on a parallel processor with a very large number of cores
(see figure \ref{f:xeon-96-machine}).
The speedup is shown in figure \ref{f:speedup-omp-bertha-molecule-size}.

Our results show a satisfactory speedup in both interaction kernels, while the 
iterative procedure again shows a poorer performance due its high memory bandwidth requirements,
which cannot be satisfied in this NUMA architecture.

\subsubsection{Hybrid MPI-thread parallelization}
\label{sss:mpi-results}

Prior to large scale computations with many Nehalem (see figure \ref{f:nehalem-machine}) nodes,
we performed test runs on one node to find an optimal OpenMP/MPI splitting.
Four calculations were done on the Nehalem machine using all 8 cores of the machine,
but differing in the OpenMP/MPI splitting. The results are collected in the table \ref{t:runtime-mpi}. 
We observed a small workload unbalance of 8\% in
case of the Hartree kernel and an even smaller unbalance of 5\% in the case of
the exchange--correlation kernel. The iterative procedure with the
``round robin distribution'' of frequencies over the nodes shows a
even lower workload unbalance of 2\%. The best run time is achieved in a 2/4 hybrid
parallelization i.e. running 2 processes with 4 cores each. This optimal
configuration reflects the structure of the machine, where each thread shares
the same L3 cache.
The node has two processors of four cores each and a relatively slower memory access between
the processors. This weak inter node communication results in an appreciable penalty
in an OpenMP-only run (1/8 configuration), because the iterative procedure reads a rather
large amount of data ($V_{\mu}^{ab}X_{b}^{E}$) twice during the application of the response function
$\chi_{0}$ (see subsection \ref{ss:matrix-vector}).

\begin{table}[htb]
\begin{center}
\begin{tabular}{|c|c||c|c|c|c|}
\hline
Proc / Thr
    & Domi. prod.
                 & $f_{\mathrm{H}}$ 
                             & $f_{\mathrm{xc}}$ & Iterative proc. 
                                                     & Total \\ \hline
1/8 & 7.3        & 271       & 142       & 145       & 571 \\ %
2/4 & 6.9 (6.9)  & 273 (267) & 142 (141) & 109 (108) & 538 \\ %
4/2 & 6.8 (6.8)  & 274 (264) & 142 (141) & 122 (112) & 544 \\ %
8/1 & 6.8 (6.8)  & 274 (257) & 143 (140) & 134 (120) & 570 \\ %
\hline 
\end{tabular}
\end{center}
\caption{Run time and speedup factors in a hybrid MPI/OpenMP parallelization
for fullerene C$_{60}$.
In the brackets, the smallest run time between the nodes is stated in order to
estimate the MPI work load disbalance.}
\label{t:runtime-mpi}
\end{table}

The high memory-bandwidth requirement is clearly revealed in two distinct ways of 
using the same number of cores (8 cores), either using all cores on one node or
distributing them over two nodes. In the latter case, the memory-bandwidth is higher
and the iterative procedure runs considerably faster (92 seconds versus 137 seconds
in the case of fullerene C$_{60}$).

\begin{figure}[htb]
\centerline{\includegraphics[width=7cm,viewport=50 60 410 300,angle=0,clip]{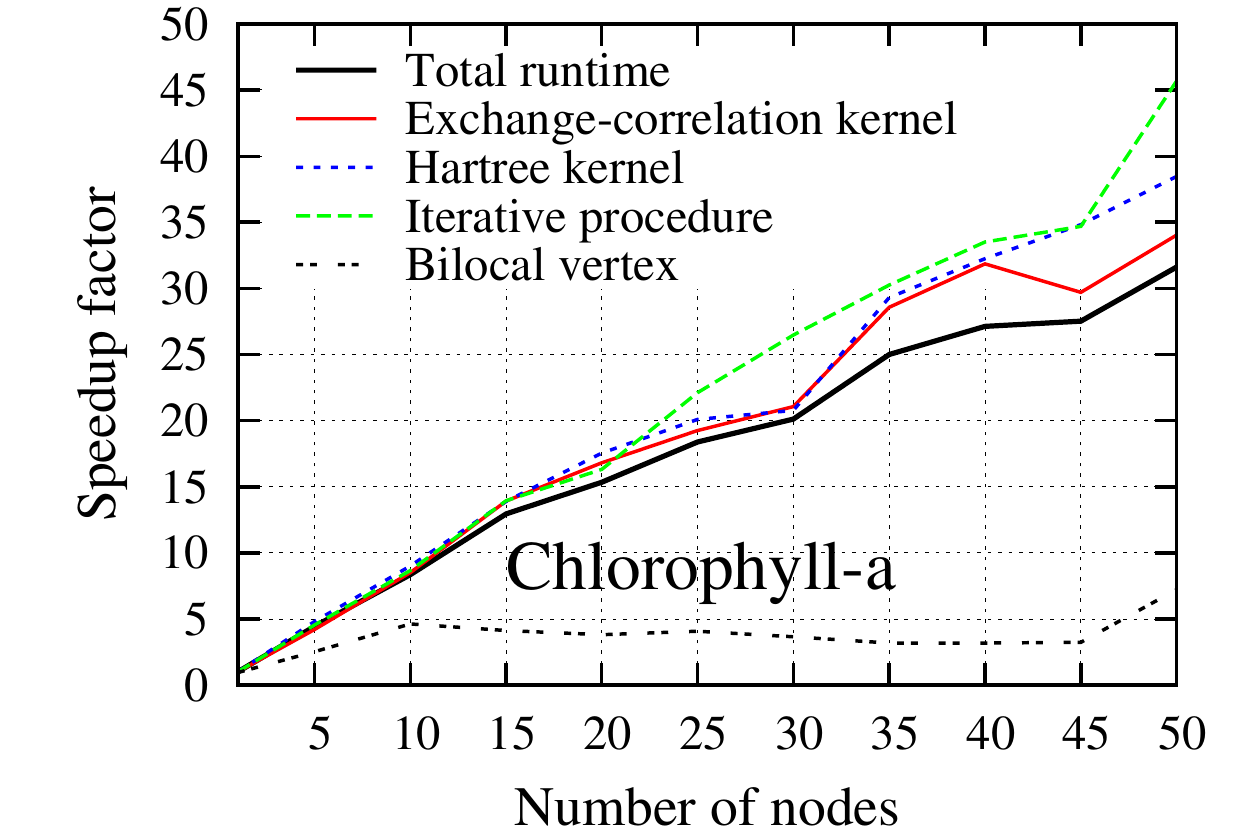}}
\caption{Speedup due to hybrid OpenMP/MPI parallelization for chlorophyll-a.
The job was run on up to 50 nodes with 2 processes per node. The code shows a linear 
speedup on up to 15 nodes (30 processes, 120 cores). Further increase of the number of nodes 
results in a steady acceleration of the whole program.}
\label{f:chlorophyll-a-speedup}
\end{figure}

We used the above optimal 2/4 hybrid configuration in a massively parallel calculation
on the chlorophyll-a molecule.
The speedup due to hybrid OpenMP--MPI parallelization is shown in figure \ref{f:chlorophyll-a-speedup}.
In this computation, we used up to 50 nodes of recent generation Nehalem machines.
According to the previous experiment on fullerene C$_{60}$, we started 2 processes per node,
each process running with four threads. The two processes were placed on sockets.
One can see that the iterative procedure shows the best speedup among other parts of the code,
while total run time is governed mainly by the calculation of the exchange-correlation kernel.
The absolute run times (including communication time) in the first calculation with one node (2 processes) were: 
total 4003 seconds, for the exchange-correlation kernel 1147 seconds,
for the Hartree kernel 1247 seconds, for the iterative procedure 1574 seconds,
and for the bilocal vertex 11.8 seconds. The speedup in the bilocal vertex reaches
a maximum at 10 nodes because of increasing communication time. 

\begin{figure}[htb]
\centerline{\includegraphics[width=7cm,angle=0,clip]{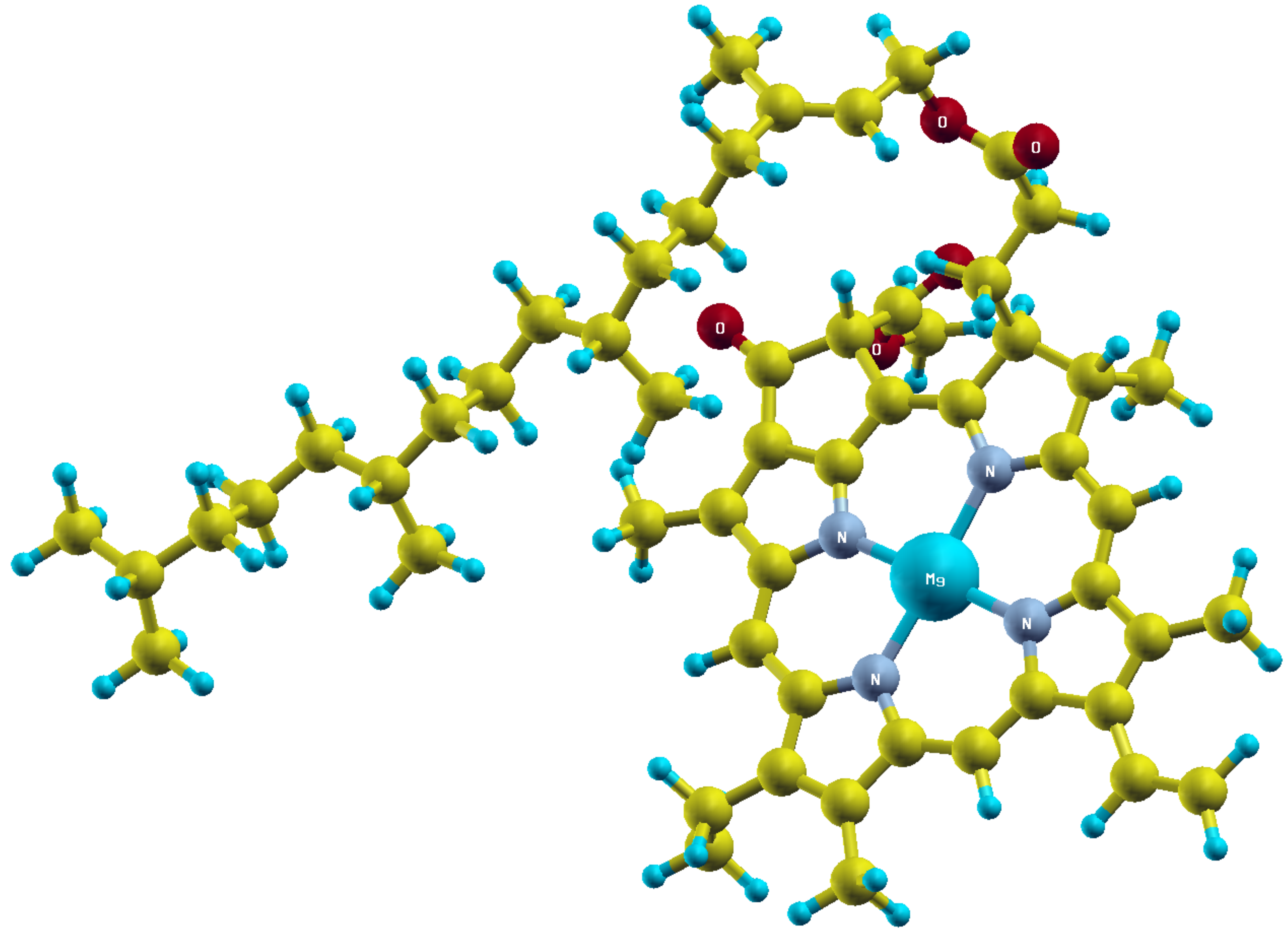}}
\caption{The relaxed geometry of chlorophyll-a molecule obtained with the SIESTA package
using a DZP basis set.}
\label{f:chlorophyll-a-geometry}
\end{figure}

\begin{figure}[htb]
\centerline{\includegraphics[width=7cm,viewport=30 60 410 300,angle=0,clip]{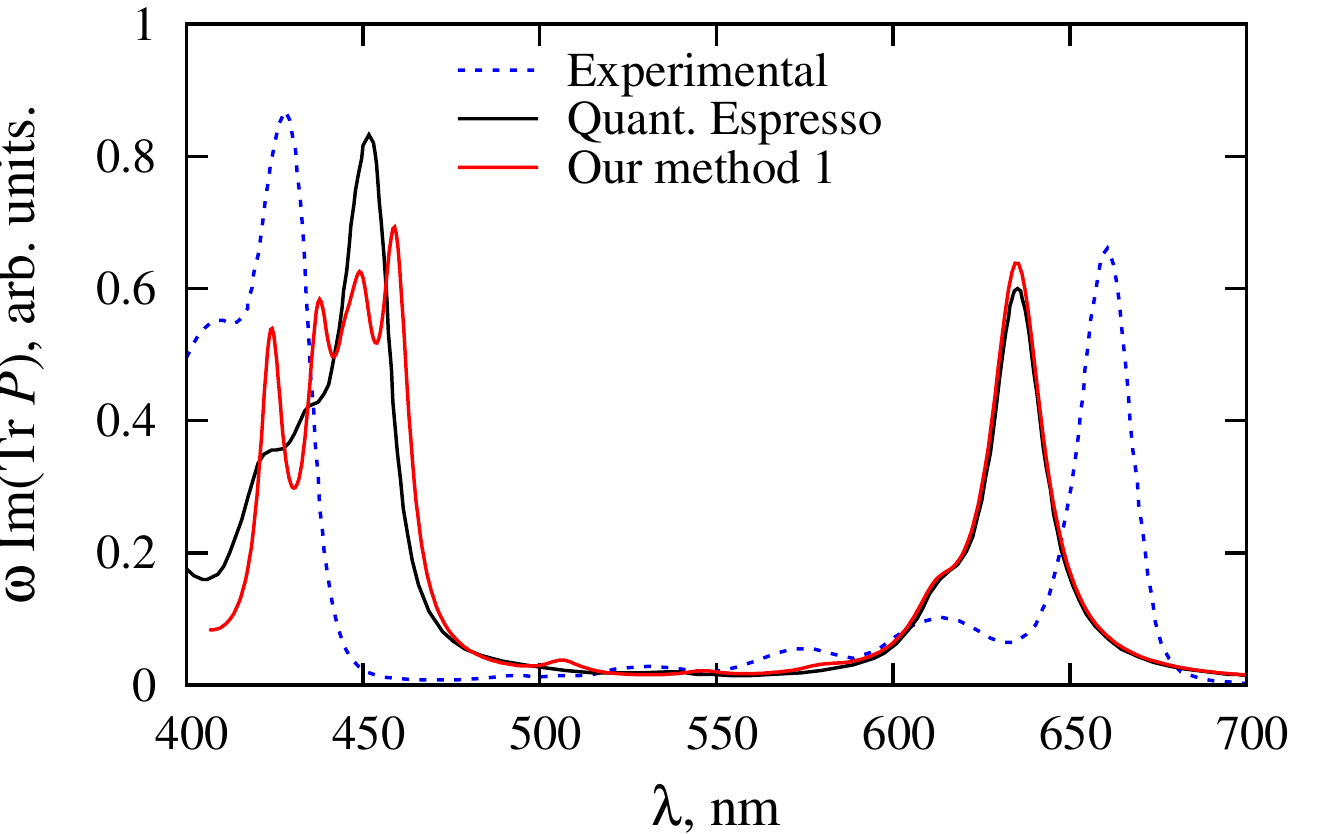}}
\caption{Low frequency absorption spectrum of chlorophyll-a.}
\label{f:chrolophyll-a-spectrum}
\end{figure}

The starting geometry of the molecule was taken from Sundholm's supplementary data \cite{Sundholm:1999}.
The geometry was further relaxed in the SIESTA package \cite{siesta} using Broyden's algorithm until the
remaining force was less than 0.04 Ry/\AA. The relaxed geometry is shown in figure \ref{f:chlorophyll-a-geometry}.
In order to achieve this (default) criterion, we had to use a finer internal mesh with a \texttt{MeshCutoff} of 185 Ry.
The default DZP basis was used, but to achieve convergence, 
we used orbitals that are more extended in space than SIESTA's default orbitals.  The spatial extension
is governed by the parameter \texttt{PAO.EnergyShift} that was set to $0.002$ Ry in the
present calculation. The spectrum of the chlorophyll-a molecule is seen in figure
\ref{f:chrolophyll-a-spectrum}. Like in the case of fullerene C$_{60}$, there
is excellent agreements between theoretical results that however differ from
the experimental data \cite{Du-etal:1998}. The low frequency spectrum consists of
two bands at  635 nm and 450 nm.
Both bands are due to transitions in the porphyrin. According to our calculation,
the first A band consist of 2 transitions between HOMO--LUMO and HOMO-1--LUMO, while the
second (so called Soret) band consist of 6 transitions to LUMO and LUMO+1 states.

\subsection{Fullerene C$_{60}$ versus PCBM}
\label{ss:fullerene-versus-pcbm}

\begin{figure}[tbp]
\centerline{\includegraphics[width=4.2cm,viewport=550 200 1050 750, angle=0,clip]{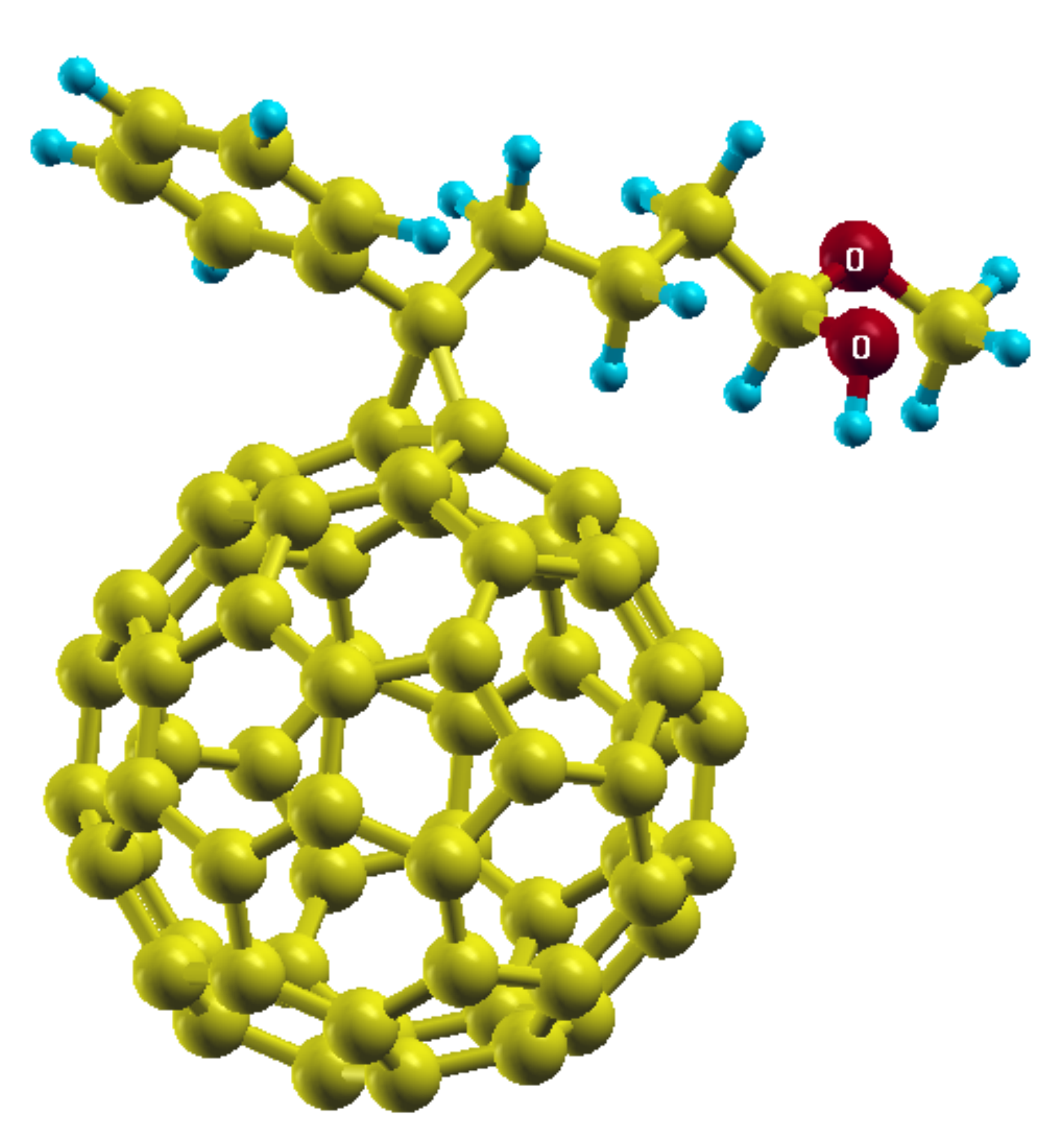}}
\caption{Geometry of PCBM. Relaxation is done in SIESTA package with
Broyden's algorithm.}
\label{f:pcbm-geom}
\end{figure}

\begin{figure}[htbp]
\centerline{\includegraphics[width=7cm,viewport=30 60 410 300, angle=0,
clip]{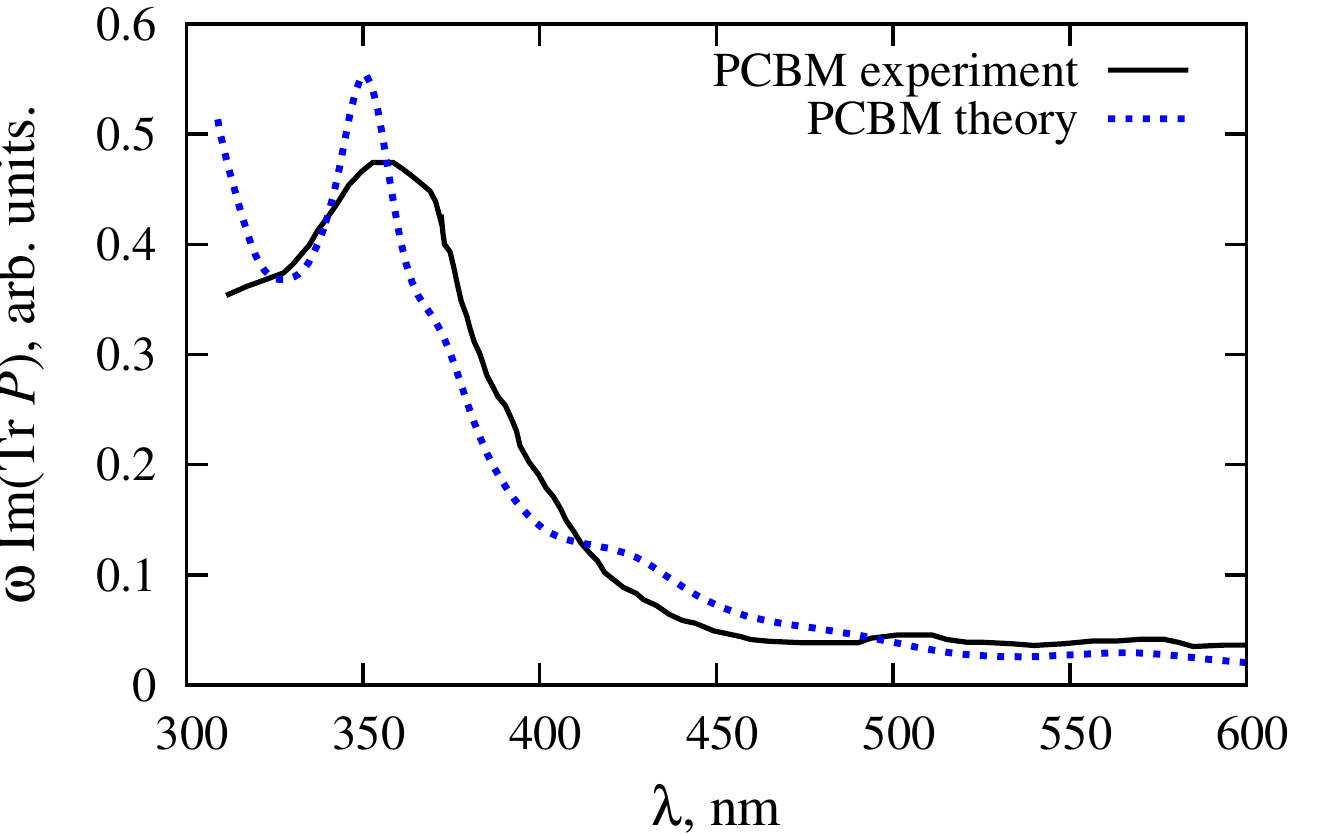}}
\caption{Comparison of the low-frequency spectra for PCBM with experimental data.}
\label{f:pcbm}
\end{figure}

\begin{figure}[htbp]
\centerline{
\includegraphics[width=7cm,viewport=30 60 410 300, angle=0, clip]{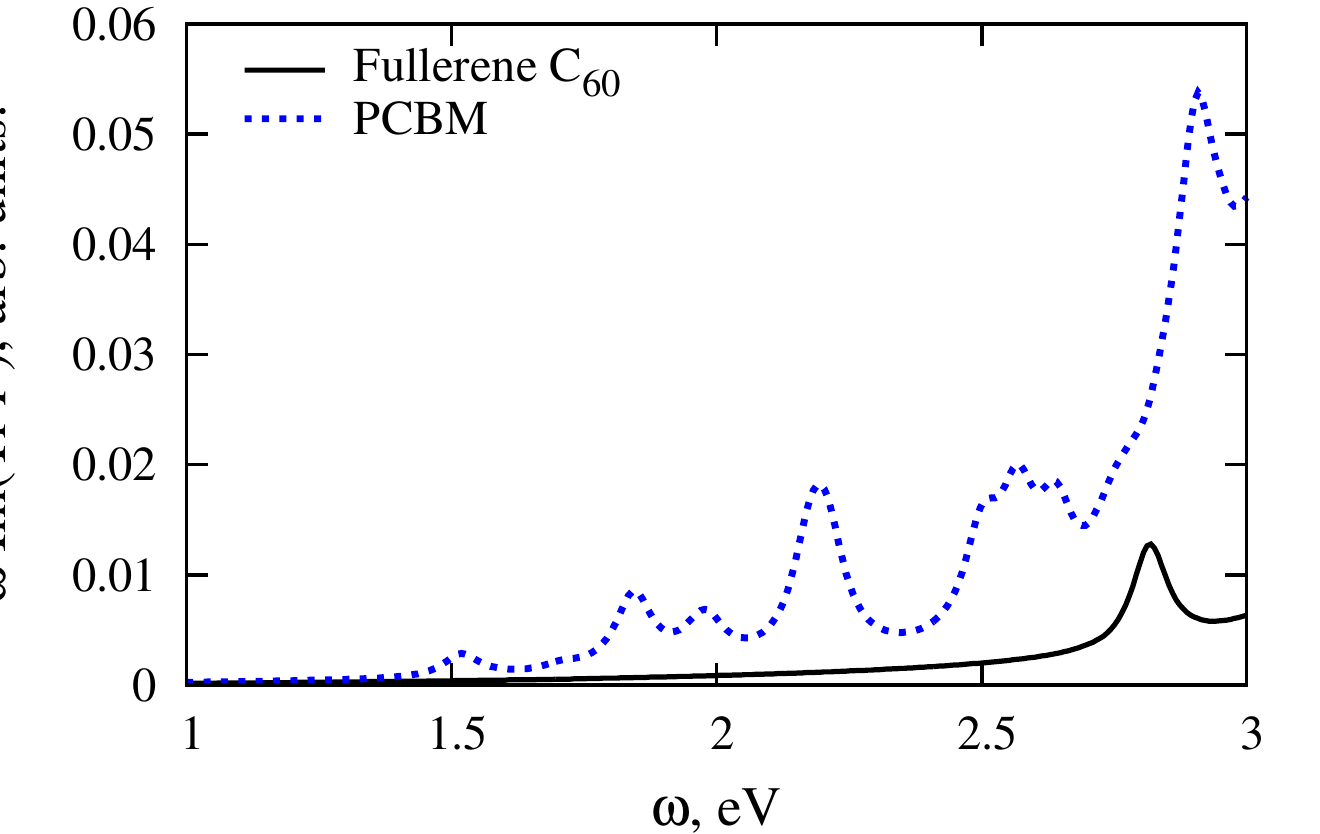}
\hspace{-0.5cm}
\includegraphics[width=7cm,viewport=30 60 410 300, angle=0, clip]{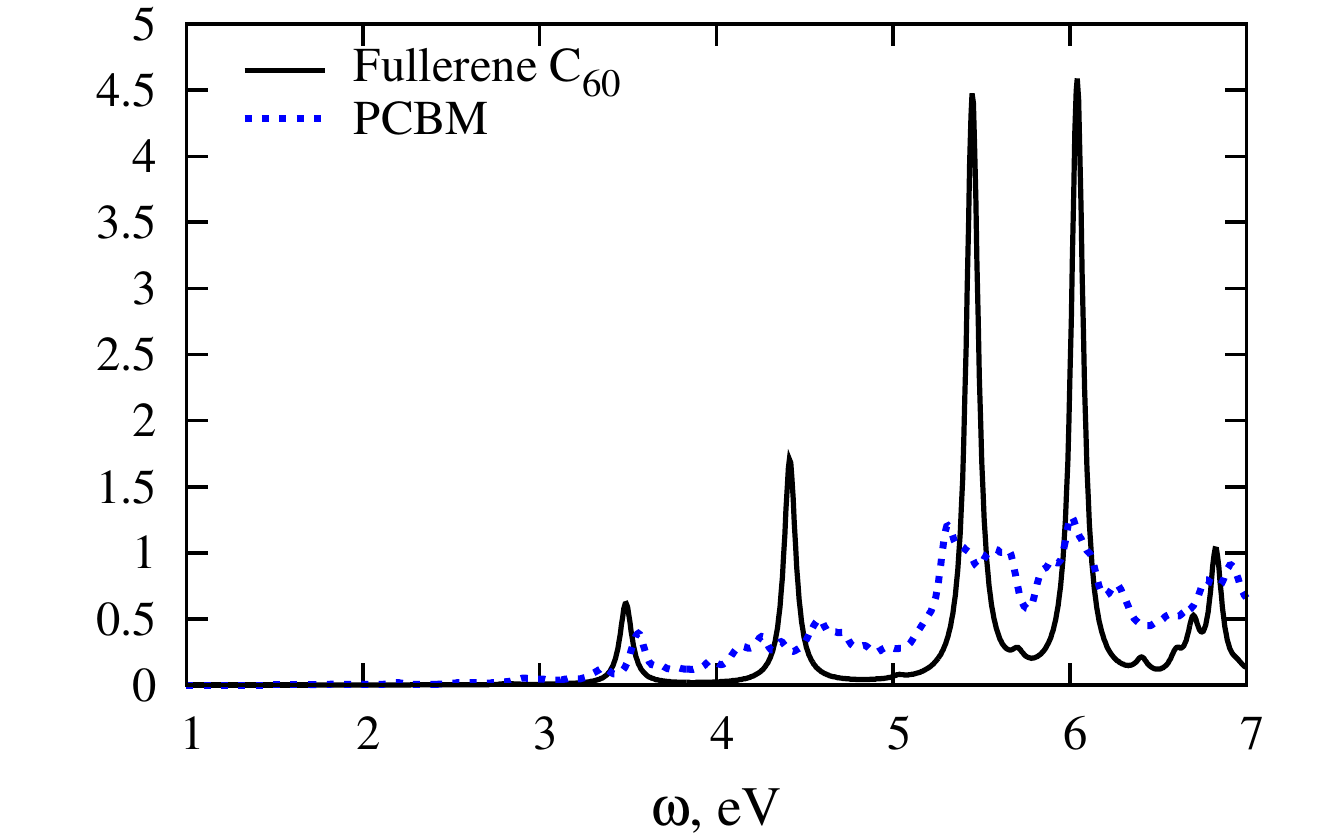}}
\caption{Comparison of the low-frequency spectra C$_{60}$ versus
PCBM.}
\label{f:pcbm-vs-c60}
\end{figure}

Fullerenes are often modified in order to tune their absorption spectra or
their transport properties \cite {Brabec:2001,Mayer:2007,Suresh-etal:2009}.
In this work, we compute the absorption spectra of [6,6]-phenyl C61 butyric acid methyl
ester (PCBM) and compare with the spectrum of pure fullerene C$_{60}$.
We use the same parameters as in the case of C$_{60}$ in subsection \ref{ss:fullerene}.
A relaxed geometry of PCBM was obtained using the SIESTA package \cite{siesta} and
using its default convergence criterion (maximal force less than 0.04 eV/\AA ).
Figure \ref{f:pcbm-geom} shows the relaxed geometry. The absorption spectrum
of PCBM is shown in figures \ref{f:pcbm} and \ref{f:pcbm-vs-c60}.

Figure \ref{f:pcbm} shows a comparison of our calculation with recent
experimental results \cite{Suresh-etal:2009}. We can see that our results
have similar features as the experimental data: the maxima at 350 nm agree
well with a broad experimental resonance at 355 nm, and a substantial
\textquotedblleft background\textquotedblright\ at longer wavelength is
present both in the calculated and in the experimental spectrum. In this calculation, we
set the damping constant $\varepsilon =0.08$ Ry and compute the spectrum in
the range where experimental data are available. However, in order to
better understand the difference introduced by the functional group, we compute
the spectra in a broader range of energies with a
smaller value of the damping constant $\varepsilon =0.003$ Ry. The result is
shown in figure \ref{f:pcbm-vs-c60}.

One can see on the left panel of figure \ref{f:pcbm-vs-c60} that PCBM
absorbs much stronger in the visible range. This is a consequence of symmetry breaking
and indicates a modified HOMO--LUMO gap. On the right panel of the
figure, one can recognize the main difference between pure and modified
fullerene. The high spatial symmetry of pure fullerene leads to a degeneracy
of the electronic transitions and several transition contribute to the same
resonance. The symmetry is broken in the case of PCBM, the degeneracy is
lifted and the spectral weight is spread out.

\section{Conclusion}
\label{s:conclusion}

In this paper, we have described a new iterative algorithm for computing
molecular spectra. The method has two key ingredients. One is a previously constructed
local basis in the space of products of atomic LCAO orbitals. The second is the computation of the
density response not in the entire space of products, but in an appropriate Krylov subspace. 

The speed of our code is roughly comparable to TDDFT codes in commercially available  software
but the reader must understand that we cannot give any details on this touchy issue. 

The algorithm was parallelized and was shown to be suitable for treating molecules of more 
than hundred atoms on large current heterogeneous architectures
using the OpenMP/MPI framework. 

Our approach leaves plenty of room for further improvements both in the method and in the algorithm.
For example, we did not consider reducing the dimension
of the space in which the response function acts, but such a reduction is feasible. 

Also, we chose a uniform mesh on the frequency axis while a more economical, adaptive choice is possible. 
We are working on an adaptive procedure to obtain good spectra with few frequency points and also to compute
the position of the poles and strength of their residues. A reduction in the  number of frequencies
will allow to avoid the full calculation of the interaction kernels,
replacing them by matrix-vector multiplications. There exist fast multipole methods  \cite{FastMultipoleMethods}
for computing fast matrix-vector products of the Hartree interaction.

Moreover, for large molecules, our embarrassingly parallel approach to compute spectra
induces a memory-bandwidth bottleneck. To avoid it, one may parallelize the frequency loop using MPI 
and parallelize the matrix-vector operations using OpenMP. 

More generally, because we do not use Casida's equations,
the methods developed here should be useful beyond the TDDFT approach,
for instance in the context of Hedin's GW approximation
\cite{HedinGW} where Casida's approach is no longer available.

\textbf{Acknowledgement}

We are indebted to Gustavo Scuseria for calling our attention to the existence of iterative
methods in TDDFT and to Stan van Gisbergen for correspondence on the iterative method implemented
in the Amsterdam Density Functional package (ADF). 

It is our special pleasure to thank James Talman (University of Western Ontario, Canada)
for contributing two crucial computer codes to this project.
We thank Luc Giraud (HiePACS, Toulouse) for discussions on the GMRES algorithm and our
colleagues Aurelian Esnard and Abdou Guermouch (University of Bordeaux) for technical advice.

We acknowledge useful correspondence on the SIESTA code by Daniel Sanchez-Portal
(DIPC, San Sebastian) and also by Andrei Postnikov (Verlaine University, Metz).
Advice by our colleagues of the ANR project CIS 2007 ``NOSSI'' especially Ross Brown and
Isabelle Baraille (IPREM, Pau), is gratefully acknowledged. We also thank Uwe Huniak (Karlsruhe,
Turbomole) for kindly supplying benchmarks of the TURBOMOLE package for comparison. 

The results and benchmarks of this paper were obtained using the PlaFRIM experimental testbed of the INRIA
PlaFRIM development project funded by LABRI, IMB, Conseil R\'egional d'Aquitaine, FeDER, Universit\'e{} de
Bordeaux and CNRS  (see \url{https://plafrim.bordeaux.inria.fr/}).

The work was done with financial support from the ANR CIS 2007 ``NOSSI'' project.

\end{document}